\documentclass[%
 reprint,
 amsmath,amssymb,
 aps,
]{revtex4-2}
\usepackage{natbib}
\usepackage{graphicx}
\usepackage{dcolumn}
\usepackage{bm}
\usepackage{color}
\usepackage[caption=false]{subfig}
\newcommand{\e}{\epsilon}

\begin{document}

\title{Bipolar Thermoelectricity in Bilayer-Graphene--Superconductor Tunnel Junctions}
\author{L. Bernazzani}
\homepage{Corresponding author: lorenzo.bernazzani@uni-konstanz.de}
\affiliation{Dipartimento di Fisica "E. Fermi", Università di Pisa, I-56127 Pisa, Italy}
\affiliation{Fachbereich Physik, Universit\"at Konstanz, D-78457 Konstanz, Germany}
\author{G. Marchegiani}
\affiliation{Quantum Research Center, Technology Innovation Institute, Abu Dhabi, P.O. Box 9639, United Arab Emirates}
\author{F. Giazotto}
\affiliation{NEST, Istituto di Nanoscienze CNR \& Scuola Normale Superiore, I-56127 Pisa, Italy}
\author{S. Roddaro}
\affiliation{Dipartimento di Fisica "E. Fermi", Università di Pisa, I-56127 Pisa, Italy}
\affiliation{NEST, Istituto di Nanoscienze CNR \& Scuola Normale Superiore, I-56127 Pisa, Italy}
\author{A. Braggio}
\homepage{https://sites.google.com/site/alessandrobraggio/Home}
\affiliation{NEST, Istituto di Nanoscienze CNR \& Scuola Normale Superiore, I-56127 Pisa, Italy.}
\date{\today}

\begin{abstract}
We investigate the thermoelectric properties of a hybrid nanodevice composed by a 2D carbon-based material and a superconductor. This system presents nonlinear bipolar thermoelectricity as induced by the spontaneous breaking of the Particle-Hole (PH) symmetry in a tunnel junction between a BiLayer Graphene (BLG) and a Bardeen-Cooper-Schrieffer (BCS) superconductor. In this scheme, the nonlinear thermoelectric effect, predicted and observed in SIS$'$ junctions, is not affected by the competitive effect of the Josephson coupling. From a fundamental perspective, the most intriguing feature of this effect is its bipolarity. The capability to open and control the BLG gap guarantees improved thermoelectric performances, that reach up to $ 1\,$mV/K regarding the Seebeck coefficient and a power density of $1~{\rm nW}/\mu{\rm m}^2$ for temperature gradients of tens of Kelvins. Furthermore, the externally controlled gating can also dope the BLG, which is otherwise intrinsically PH symmetric, giving us the opportunity to investigate the bipolar thermoelectricity even in presence of a controlled suppression of the PH symmetry. The predicted robustness of this system could foster further experimental investigations and applications in the near future, thanks to the available techniques of nano-fabrication.
\end{abstract}

\maketitle

\section{Introduction}

Thermoelectricity is conventionally investigated in materials where there is a well-defined dominant carrier \cite{Tritt,TrittSub,Barber}. Indeed, the Boltzmann theory of transport demonstrates that in the linear response regime the Seebeck coefficient has the same sign of the charge of the dominant carrier \cite{Tritt,Grosso}. In other words, thermoelectricity seems to be present only in systems where the PH symmetry is broken to a certain extent \cite{Benenti}.
This becomes paradigmatic, in superconducting systems, where it should be expected that PH symmetry and the dissipationless flow of Cooper pairs would implicitly limit any thermoelectric effect. 
Notwithstanding, a few works \cite{Ginzburg,SmithTinkham} anticipated that even superconducting systems could exhibit a thermoactive behaviour. Along this route, theoretical \cite{Machon13,Ozaeta14,SanchezML2,SanchezML1,Pershoguba19,Virtanen04,Jacquod10,Kalenkov17,Hussein19,Kirsanov19,Das21} and experimental~\cite{Eom98,Jiang05,Tan21} research has shown that nonlocality or phase coherence may still trigger intriguing thermoelectric phenomena in superconducting and hybrid systems. Furthermore, some authors even reported the existence of Absolute Negative Conductance (ANC) in superconducting tunnel junctions in out-of-equilibrium conditions or under microwave irradiation \cite{Spivak,Gershenzon1,Gershenzon2,Flokstra}.\par
More recently, instead, in a number of theoretical papers~\cite{MarBraGia1,MarBraGia2,MarBraGia3,MarBraGia4}, the spontaneous breaking of the PH symmetry was predicted in Superconductor-Insulator-Superconductor' (SIS$'$) tunnel junctions subject to a finite thermal gradient. In these works, a strong bipolar thermoelectric effect was clearly identified in the nonlinear regime of the thermal bias. These latter predictions were confirmed experimentally~\cite{GaiaNature,GaiaPRApp} providing also with methods to make superconducting non-volatile memories~\cite{MarBraGia1,Patent} or other spin-active devices~\cite{Germanese2021}.\par

However, there are several drawbacks concerning the physical realisation of the cited effect in asymmetric SIS$'$ junctions. The most serious issue is given by the Josephson coupling, which needs to be substantially suppressed~\cite{GaiaNature}, as theoretically anticipated in Ref. \onlinecite{MarBraGia3}. Indeed the Josephson effect may completely kill the thermoelectric generation of power by shunting the junction at low biases, i.e. $|eV|\ll(\Delta_R+\Delta_L)$, here $\Delta_{R/L}$ are the values of the bulk superconducting gap function on the two sides of the SIS$'$ junction, therefore restoring the PH symmetry and suppressing the thermoelectric effect. This limitation may be even more serious in out-of-equilibrium conditions since the Josephson coupling presents anomalous jumps when thermal gradients are applied to asymmetric SIS$'$ superconducting  junction \cite{Guarcello1,Guarcello2}.
Hence, the possibility of making a junction using instead a semiconductor coupled to superconductors would be beneficial in this respect. Furthermore, this nonlinear thermoelectric effect manifests strongly at the so-called "matching peak" conditions $V_{peak}=\pm|(\Delta_R-\Delta_L)/e|$, where the singularities in the BCS DoS are aligned. So the possibility to engineer in detail the semiconducting gaps would be beneficial for the performance of the device.\par

Nevertheless, the aforementioned bipolar thermoelectric effect requires that the semiconductors forming the junction have a gap comparable with those of the superconductors which they are coupled to, i.e. they should have gaps of the order of few meV \cite{KittelSS,Tinkham}. On the other hand, elemental or binary compound semiconductors have gaps larger than hundreds of meV \cite{KittelSS}. This fact pushed us to consider a completely different platform, i.e. carbon-based low dimensional materials, that can exhibit smaller bandgaps, whose value can even be tuned in some cases by electrostatic gating \cite{UniPiNonlinear}.\par

Hereafter we demonstrate the bipolar thermoelectricity in hybrid junctions composed of BiLayer Graphene (BLG) coupled to BCS superconductors (S) through an insulating barrier (I). Notably, the bandgap of BLG is controllable by electrical gating \cite{NatureMaterials}, making possible to adapt the bandgap to the best operating conditions, which are connected with the superconducting gap and with the temperature of the hot lead (BLG), by changing just the electric gates voltages. This peculiar phenomenon increases the flexibility and the performances of the device. For instance, it allows one to switch on and off thermoelectricity in the junction.
Moreover, the replacement of BLG in place of one superconducting electrode relaxes also the constraints on the hot electrode temperature, leading to an even wider applicability of the thermoactive elements, which in principle could operate up to hundred kelvins.\par

\section{Device description}
\begin{figure}[htp]
 \centering
     \subfloat{\includegraphics[scale=0.17]{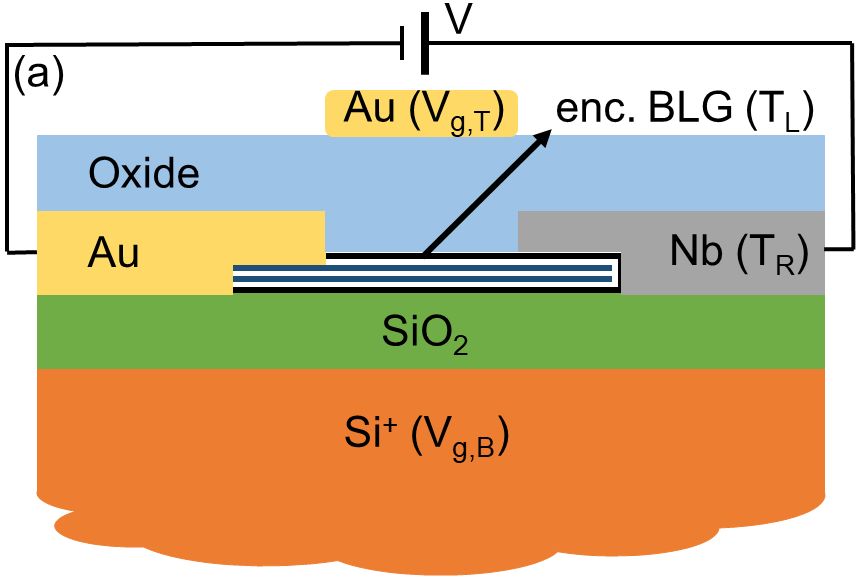}\label{<figure1>}}\,
     \subfloat{\includegraphics[scale=0.19]{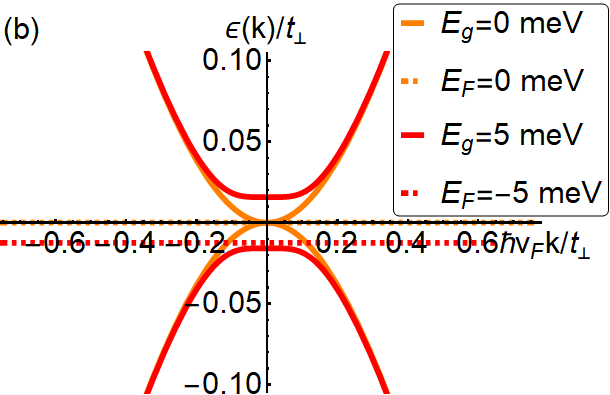}\label{<figure2>}}
     \caption{
     a) Schematics of the proposed device. The BiLayer Graphene (BLG) is placed on top of a Silica (SiO$_2$) dielectric ($\varepsilon_b=3.9$) 100-nm-thick, beneath which there is the bottom gate (doped Si). BLG is encapsulated in a hBN monolayer but is Ohmic coupled with a Au electrode on the left side. The Niobium (Nb) superconducting electrode is deposited partially on top of the BLG. hBN constitutes the tunnelling barrier with the superconducting electrode. Finally, an oxide ($\varepsilon_t=7.5$)\cite{NatureZhang}, approximately 190-nm-thick, is deposited above the electrodes and a top-gate made from Au is placed on top of this dielectric. b) Band scheme of the BLG \cite{Castro1,McCann1,NoteNew}. A gap has been opened using the top/bottom gates. With the dashed lines we also show how the BLG Fermi energy $E_F^{BLG}$ moves along the bands, due to an asymmetric gating effect. We assumed $v_F\approx 10^6\,\rm m/s$ as the Fermi velocity in graphene.}
     \label{Dis}
\end{figure}
Low dimensional systems have attracted a lot of interest recently \cite{Baladin} and intriguingly have been already discussed as promising materials for improving thermoelectric performance in the pioneering work of Ref. \onlinecite{Dresselhaus1999}. BLG, in particular, is a two-dimensional material formed by two planes of graphene stacked onto each other. It naturally inherits the intrinsic PH symmetry of graphene if it is not extrinsecally doped. However its bands deviate from the conventional linear dispersion relations of graphene \cite{Novoselov2005,Sprinkle2009}. In fact, by tight-binding computations \cite{McCann2,Slavinska} one can show that they are characterised by a nonlinear dispersion relation \cite{Mak}.
Similarly to monolayer graphene, BLG is a semimetal, and, in the pristine form, valence and conduction bands touch each other at the high symmetry K, K' points. Notably, by applying an electric field perpendicular to the BLG plane, a gapped phase can be induced, effectively behaving as an electrically controllable semiconductor. In Fig.~\ref{Dis}a we display a sketch of the setup discussed below, where the BLG is coupled to a top and a bottom gate that may be tuned externally. The gate-controlled resultant bandgap is highlighted in Fig.~\ref{Dis}b. In the absence of gating, the BLG is gapless  and the dispersion of the lowest-energy band is approximately quadratic $\e(k)=\pm (\sqrt{1+4(v_F\hbar k)^2}-1)/2\approx \pm(v_F\hbar k)^2$
(orange solid curves)~\cite{Castro1}, with $v_F$ the Fermi velocity.
BLG's gap can be opened and controlled by applying a differential gating between the backgate $V_{g,B}$ and the topgate $V_{g,T}$ \cite{Castro1,McCann1} (solid red curves). However, gating may also induce charge doping at the BLG planes, spoiling the perfect PH symmetry and paving the way to the intriguing possibility of investigating a bipolar thermoelectric device in the presence of a controlled breaking of the PH symmetry as well. As reported in \cite{McCann1} this tunable bandgap can be computed using an approximate formula, according to which it is given by \cite{Castro1,McCann1},
\begin{equation}
\label{BLGGap}
2E_g=\frac{|U|t_\perp}{\sqrt{U^2+t_\perp^2}}\,,
\end{equation}
where $U$ is the potential energy difference between the two graphene planes, also called the gating asymmetry parameter, and $t_\perp\approx0.4\,$eV is the inter-layer hopping energy term in the tight binding approximation of BLG's electronic bandstructure. The approximate energy dispersion in the gapped state~\cite{Castro1} $\e(k)\approx \pm\sqrt{U^2/4+\pm(v_F\hbar k)^4/t_\perp^2}$ is shown in Fig.~\ref{BLGGap}b (solid red curve).
\par
This paper deals with the theoretical study of a  BLG-I-S tunnel junction, whose schematic representation is given in Fig.~\ref{Dis}a. The BLG is on the left side of the junction and is deposited on the Silica substrate. We suppose to be able to heat up the BLG electronic temperature in order to induce an effective thermal gradient across the junction by means of irradiation by a photon source or by the Joule effect due to normal currents flowing between electrodes \cite{Vischi,Viljas} (this can be usually accomplished using other normal metal electrodes, e.g. yellow contact in Fig.~\ref{Dis}a).
The BLG is in clean contact with a gold (Au) electrode (see Fig.~\ref{<figure1>}), so that the voltage drop $V$ applied across the structure localizes at the BLG-I-S junction. In other words, the Au electrode and the BLG are equipotentials.
The electrostatic gating is applied with the aid of a top-gate $V_{g,T}$ capacitively-coupled to BLG through an oxide and a $Si^+$ layer, with the role of the bottom gate, beneath the Silica layer, as shown in Fig.~\ref{Dis}a \cite{NatureZhang}\footnote{The top/bottom gates are referred to the BLG Ohmic contacts in order to induce a proper gating on the BLG which is not affected by the bias applied to the junction.}.
The right electrode composing the junction is a Bardeen-Cooper-Schrieffer (BCS) superconductor (grey in Fig.~\ref{Dis}a). In order to have a superconducting gap commensurate with those of BLG and, at the same time, a good range of operating temperatures, we chose Nb for the BCS superconductor. This metal has a critical temperature of $T_c\approx 9\,\rm K$. It is deposited partially on the encapsulated BLG in order to provide a tunnel contact with it. The encapsulation of BLG is made from hexagonal Boron Nitride (hBN) monolayer. This material constitutes proper tunnel barriers since it is a good insulator with an almost perfect crystallographic order \cite{Maestre2021}.
\par

\section{Bipolar thermoelectricity with PH symmetry}
\label{sec:PHsymmetric}
In the tunnelling framework, we can compute the charge and heat currents according to the following formula \cite{Tinkham1972,Tinkham,Ketterson,MarBraGia1,MarBraGia2}
\begin{align}
\label{Eq:IQ}
\begin{bmatrix}
I   \\   
\dot Q_{L}
\end{bmatrix}
=\frac{4\pi|\mathcal{T}|^2}{\hbar}\int_{-\infty}^{+\infty}d\varepsilon& 
\begin{bmatrix}
-e \\
\varepsilon
\end{bmatrix} N_{BLG}(\varepsilon)N_{S}(\varepsilon+eV)\cdot\nonumber\\
&\cdot[f_L(\varepsilon)-f_R(\varepsilon+eV)]
\end{align}
where $|\mathcal{T}|$ is the transmittivity of the tunnelling barrier, which in our case is given by the hBN layer between the BLG and the superconducting electrode. In writing Eq.~\eqref{Eq:IQ}, we further assumed that the barrier transmissivity is independent of the gate voltages $V_{g,T}$ and $V_{g,B}$ applied to the BLG. This approximation is reasonable at the energy scales considered in this work, where the induced BLG bandgaps and the applied voltage bias are the order of few meV, but may fail for a very strong gating. The electrons' gas in the semi- and superconductive leads follows the Fermi-Dirac distributions $f_i(\epsilon,T_i)=[1+\exp{(\epsilon/k_BT_i)}]^{-1}$, where we assume that carriers are in quasi-equilibrium, being well described by a standard quasiparticle distribution with an electronic temperature $T_i$ \cite{RMPGiazotto}. We introduce also the tunnelling barrier bias voltage $V$.\par
In Eq.~\ref{Eq:IQ} the Density of States (DoS) of the two electrodes are
$N_{BLG/S}(E)$. In our case the right electrode is a BCS superconductor, therefore 
\begin{equation}
N_{S}(\varepsilon)=N_F\,\Biggl|\Re{\Biggl[ \frac{\varepsilon+i\Gamma_D}{\sqrt{(\varepsilon+i\Gamma_D)^2-\Delta_R(T_R)^2}}\Biggr]\Biggr|}
\end{equation}
with $N_F$ the DoS of the metal in the normal state at the Fermi energy, $\Delta_R(T_R)$ is the superconducting gap and $\Gamma_D$  is the Dynes' phenomenological correction to BCS DoS \cite{Dynes}. On the other hand, the left electrode is composed by the BLG whose DoS for $E\ll t$ can be approximated with the following expression \cite{Castro1,Suarez}:
\begin{equation}
\label{Eq:BLGDOS}
N_{BLG}(\epsilon)\approx\frac{t_{\perp}}{\sqrt{3}\,\pi\,t^2}\cdot\frac{|E|\,\Theta(|E|-E_g)}{\sqrt{{E}^{2}-E_g^{2}}}
\end{equation}
where $\Theta(x)$ is the Heaviside step-function and $E=\epsilon+E_{F}^{BLG}$ with $E_{F}^{BLG}$ being the BLG's Fermi energy. The tight-binding energies for the in-plane hopping integral of the graphene is $t\approx 3\,$eV and the out-of-plane hopping term is $t_{\perp}\approx 0.4\,$eV.
We assume the insulating barrier between BLG and superconducting electrodes being the hBN
\footnote{The hBN is commonly used to encapsulate graphene since its crystalline configuration matches well with that of graphene
diminishing the surface disorder at the interfaces.}. We also assume to have a tunnelling conductance $G_T\approx (10\,{\rm k}\Omega)^{-1}$ \footnote{The conductance $G_T=4 N_Fe^2|\mathcal{T}|^2t_\perp/(\hbar\sqrt{3}t^2)$ would represent the differential conductance of the tunnel junction when the superconductor is in the normal state and the BLG is gapless.} for the hBN separating the BLG electrode, above which the superconducting electrodes will be deposited forming a junction of about $0.1\,\mu \rm{m}^2$, this choice is consistent with values reported in Ref. \onlinecite{NanoLetters}. Despite a larger active surface would lead to a higher conductance that would be beneficial for the performance of the proposed junction, it is nonetheless important that the barrier between the leads is a good insulating barrier, since we are going to establish a sizeable temperature gradient across the junction. Furthermore, enlargement of the active surface could also lead to unwanted effects such as pinhole points \cite{PinHole}.\par
Clearly, either asymmetric gating or charge traps at the oxide surface may also induce a deviation in the midgap position of the BLG's Fermi energy $E_F^{BLG}$ partially breaking the PH symmetry. However, with two independent top and bottom gates it is always possible to restore the PH symmetry in the BLG. Indeed, since the bandgap $2E_g\approx U$ (for $U\ll t_{\perp}$, see Eq.~\eqref{BLGGap}) is induced mainly by the differential gating mode, the common mode can be exploited to modify $E_F^{BLG}$ changing the BLG doping \cite{Castro1,McCann1,Padilha,Castro2,McCann2,NatureZhang}. More specifically, we can find a line of points in the $(V_{g,B},V_{g,T})$ plane for which the BLG turns out to be fully PH symmetric and gapped. This is a straight line with a slope of $-C_{b}/C_{t}$
where $C_{i}=\varepsilon_0 \varepsilon_{i}/d_i$ are the capacitances with top ($i=t$) or bottom  ($i=b$) gates determined by the thicknesses $d_b$($d_t$) oxide layers and their dielectric constant $\varepsilon_t$ ($\varepsilon_b$) \cite{NatureZhang}. We remark that the charge accumulated in the BLG through gating  modifies the Fermi energy and so the total junction differential conductance (see discussion in Sec.~\ref{sec:PHsymmetric}). However, these effects cannot be interpreted as a gating dependence of the transmissivity of the tunnelling barrier.
Therefore, in the first part of this paper, we investigate the device in the gating conditions where the BLG electrode is fully PH symmetric, i.e. $E_F^{BLG}=0$ according to our notation. In this configuration, due to the PH symmetric nature of the electrodes, one should expect that the junction could never display a linear thermoelectric behaviour \cite{Benenti,DiVentra}. Notwithstanding, when BLG's electronic temperature and the thermal gradients exceed finite thresholds, we predict the possibility to induce spontaneously a thermoelectric regime in the junction just by tuning BLG's gap at values higher than Nb's superconducting gap.
Furthermore, it is easy to check that the PH symmetry of the two leads, i.e. $N_i(E)=N_i(-E)$, would result in the anti-symmetric nature of the current with respect to the electric bias, i.e.
$I(V)=-I(-V)$. This phenomenology is similar to that reported in \cite{MarBraGia1,MarBraGia2,GaiaNature}. However, it is worth noting that the possibility to tune the electrode gap through gating, is quite unique to this system. In fact in the SIS$'$ case one should change the material composing the electrode \cite{GaiaNature} or recurring to complex phase-controlled gaps in proximity junctions \cite{Guarcello2}.\par

\begin{figure}
\centering
     \subfloat{\includegraphics[scale=0.2]{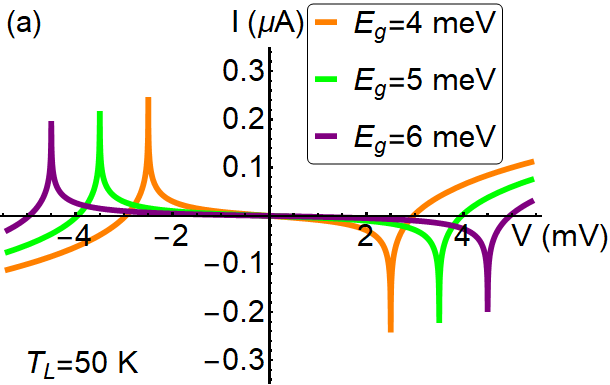}}
     \subfloat{\includegraphics[scale=0.2]{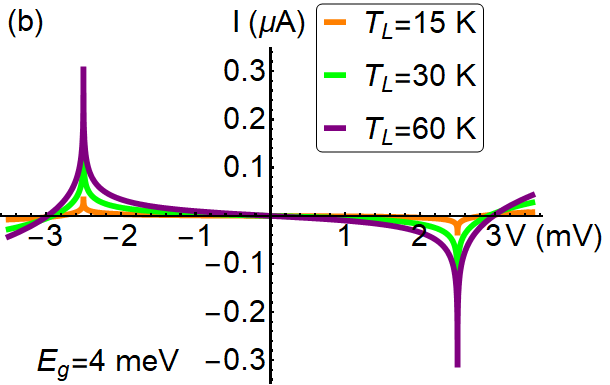}}
     \caption{
     IV characteristics of the proposed junction in the thermoelectric regime $E_g>\Delta_R(T_R)$. The thermoelectricity is easily spotted because $IV<0$ and this implies ANC. a) IV characteristics plotted by varying the BLG's gap (that is obtained by varying the gating voltages while keeping the BLG in the PH symmetry line). Here the BLG temperature is kept at $T_L=50\,$K b) IV characteristics plotted varying the electronic temperature of the BLG while keeping the BLG gap fixed at $2E_g=8\,$meV. The height of the peak, and consequently the strength of the thermoelectricity, grows by increasing the hot lead temperature. Quite peculiarly, due to the nonlinearity of the effect, it appears that the Seebeck voltage $V_S$, defined by $I(V_S)=0$, depends weakly on the BLG temperature being more connected to the position of the matching peak $V_p$ instead, that in this system is unaffected by the $T_L$. In the computation we assumed the superconductor to reside at the bath temperature of $T_R=T_{bath}=1\,$K, a Dynes parameter $\Gamma_D/\Delta=10^{-4}$, and tunneling conductance $G_T=(10~{\rm k}\Omega)^{-1}$.}
     \label{CharPH}
\end{figure}

The discussed bipolar thermoelectric effect is nonlinear since a finite temperature difference is required to activate the junction. When $E_g>\Delta_R(T_R)$ the junction can be in the thermoactive regime, i.e., displaying ANC $IV<0$, for an appropriate temperature difference between the two terminals. In Fig.~\ref{CharPH} we show the IV characteristics in the subgap regime when the junction is thermoactive. The electronic temperature of the BLG is assumed to be kept at a high temperature (e.g. $T_L=50\,$K in Fig. \ref{CharPH}a) and the superconductor will be on the cold side with $T_R=T_{bath}=1\, $K \footnote{In principle inverting the ratio between $E_g/\Delta_{0,R}$ also the opposite temperature gradient may be assumed. But this is less practical since the temperature of the superconducting lead should not exceed its critical temperature in any case.}. Similarly to the SIS$'$ case~\cite{MarBraGia2}, we can identify two different thermoelectric regimes that depend exclusively on the absolute module $|V|$. These are the linear-in-bias regime, that is revealed by the negative differential conductance at $V\approx 0$, and the nonlinear-in-bias effect which manifests strongly with the biggest ANC at the matching peak condition $|eV_p|=E_g(U)-\Delta_R(T_{bath})$.
By looking at the peaks in Fig. \ref{CharPH}a,b,
we observe how their position is changed following the evolution of the BLG gap $E_g$ that is modified by external gating. Notably, we see that the height of the ANC peaks is weakly dependent on the BLG gap (see \ref{CharPH}a) but is more affected by the temperature of BLG (see \ref{CharPH}b). Finally, we note that in this configuration where the BLG is hot and the superconductor is cold, the hot temperature does not affect the position of the peak since the $E_g$ depends only on the electrostatic potential $U$. This is a substantial difference with respect to the SIS$'$ case \cite{MarBraGia1,MarBraGia2,GaiaNature}. As a consequence, it is also important that the thermoelectric power $P=-IV$ is a monotonously increasing function of the hot temperature in our system in contrast to the SIS$'$ concept where the closing of the gap with temperature limits the thermoelectricity to a well-defined range, spanning at most an interval of few Kelvins.\par

\begin{figure}
     \subfloat{\includegraphics[scale=0.25]{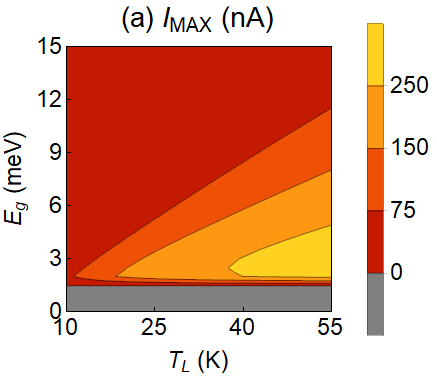}\label{<figure1>}}\,
     \subfloat{\includegraphics[scale=0.25]{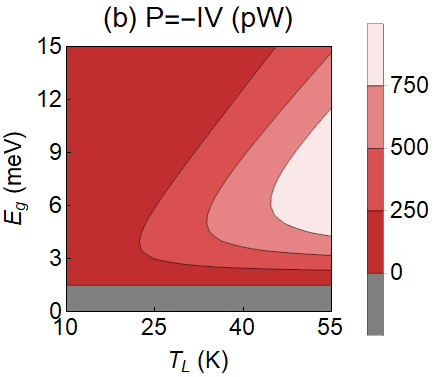}\label{<figure2>}}\\
     \subfloat{\includegraphics[scale=0.25]{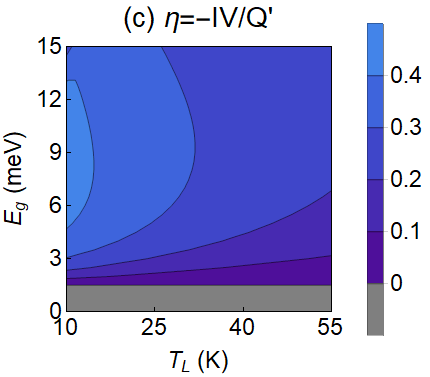}\label{<figure2>}}\,
     \subfloat{\includegraphics[scale=0.25]{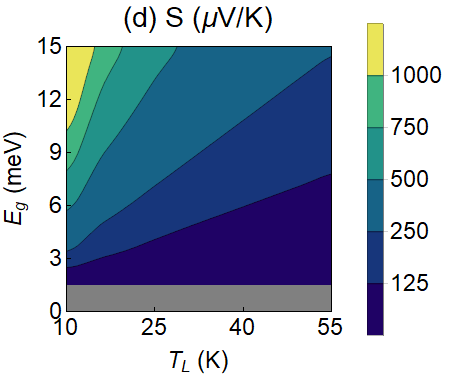}\label{<figure2>}}
     \caption{
     Thermoelectric figures of merits for the proposed junction at the PH symmetric point as a function of the BLG bandgap $2E_g$ and the BLG electronic temperature $T_L$. a) Displays the maximum current in nA obtained at the matching peak condition $I(V_p)$. b) Shows the thermoelectric power at the matching peak $-I(V_p)V_p$ in units of pW. c) The thermodynamic absolute efficiency computed at the matching peak $V_p$, where the output power is roughly maximum. d) Displays the nonlinear Seebeck coefficient in units of $\mu \rm V/K$. The remaining parameters are the same of Fig.~\ref{CharPH}.}
     \label{WP}
\end{figure}

Since we are not dealing with a linear thermoelectric effect, we cannot rely on standard figures of merit typical for thermoelectric elements, such as ZT \cite{Benenti,Splett}. Hereafter we will focus on alternative figures of merit in order to quantify the performances of the BLG-I-S thermoelectric junction in comparison with other recent proposals \cite{GaiaNature}. In particular, it is interesting to show them as in Fig.~\ref{WP} by using contour plots. Therefore investigating how they depend on both the BLG gap $E_g$ and temperature $T_L$. In Fig.~\ref{WP}a we show the maximum thermocurrent $I_{M\!A\!X}=|I(\pm V_p)|$, which is obtained at the matching peak voltage $V_p$ and, correspondingly, in Fig.~\ref{WP}b the  thermoelectric power delivered at the same point $P(V_p)= -I(V_p)\cdot V_p$ by the junction \footnote{Even if the absolute value of those quantity depends also on the Dynes' parameter $\Gamma_D$ the general behaviour would be quite independent on it.}.
The grey areas in the plots denote the regions with $E_g<\Delta_R(T_R)$, where the system is purely dissipative, thus producing no net thermoelectric current.
Both of the quantities are monotonically increasing functions of the BLG temperature $T_L$ and quickly saturate for $T_L\gtrsim E_g/k_B$ (not shown).  These figures of merit behave not monotonically with respect to the BLG gap $E_g$, displaying a maximum for a temperature-dependent optimal value. We notice that the values of the thermoelectric charge current are quite high for a junction of $0.1\,\mu\rm m^2\,$. At the same time the output power, shown in figure \ref{WP}b, is notably high, since it reaches values above $750\,$pW for high-temperature gradients and intermediate gap values which correspond to a power density of roughly $1\,\rm{nW}/\mu\rm{m}^2$. Since it is computed at the matching peak, it is natural to ask how this quantity would be affected by the quality of the superconducting contact, leading to a higher Dynes' parameter. We report, without showing it, that the output power of the junctions is only reduced by a logarithmic correction of about a factor of 2 when changing the Dynes' parameter by two orders of magnitudes (from the $10^{-4}\Delta_0$ used here to $10^{-2}\Delta_0$).\par

Furthermore, it is interesting to discuss, looking at Fig.~\ref{WP}c, the absolute efficiency of the thermoelectric engine at the maximum power. An useful approximation for such quantity is the value it assumes at $V=V_p$, i.e. $\eta_p\approx-I(V_p)V_p/\dot{Q}_L(V_p)$, since the junction delivers maximum power approximately at the matching peak point. At the same time, the heat losses in the junction is probably dominated by the quasiparticle contribution $\dot{Q}_L(V_p)$ computed using Eq.~\ref{Eq:IQ} \footnote{Obviously the heat may potentially be transferred  also by other contributions such as phonon or even radiative  contributions that are more difficult to be estimated. Our computation of efficiency represents a sort of optimal electronic efficiency of the device.}.
In this case the best efficiencies are obtained at the lowest BLG temperatures $T_L$, where the decrease in the heat losses is much bigger than the decrease of the thermopower. Notably, the absolute efficiency reaches values up to $40\%$ and, when the hot temperature is fixed, there is an optimal value for the gap.
It is worth noting that the efficiency here is the absolute thermodynamic efficiency of the thermodynamic machine.
\par
Finally, in Fig. \ref{WP}d the thermoelectric performance, is shown in terms of Nonlinear Seebeck Coefficient (NSC) which is computed as \begin{equation}
S=\frac{V_S}{T_L-T_R}
\end{equation}
\cite{MarBraGia1}, where the $V_S$ is computed numerically by finding the point at which the IV characteristic crosses the zero current axis, i.e. $I(V_S)=0$. This is a natural extension, in a nonlinear picture, of the Seebeck coefficient defined as $S=\lim_{\Delta T\to 0}\frac{\Delta V}{\Delta T}\bigl|_{I=0}$. By the way, that is the solution of the linear thermoelectricity equation $I=G\,\Delta V - G\,S\,\Delta T=0$ \cite{Benenti}.
This quantity can be computed only in the region where the junction is thermoelectric ($IV<0$), while the dissipative regions are coloured in grey. 
The values of the NSC demonstrate the strength of the thermoelectricity generation in this junction, that reaches values higher than $1000\,\mu\rm V/K$ at the lowest temperature gradient explored of $\approx 9\,$K. This graph reveals also a quite general behaviour of this quantity which increases with the BLG bandgap. Looking at panel \ref{WP}d we can say that the highest values of the NSC can be reached at low values of the temperature gradient where the thermopower is quite low (see Fig.~\ref{WP}b) confirming once again, even in this nonlinear situation, a sort of trade-off between efficiency and power production.

\section{Partial breaking of PH symmetry}
\label{sec:Breaking}
\begin{figure*}
 \centering
     \subfloat{\includegraphics[scale=0.25]{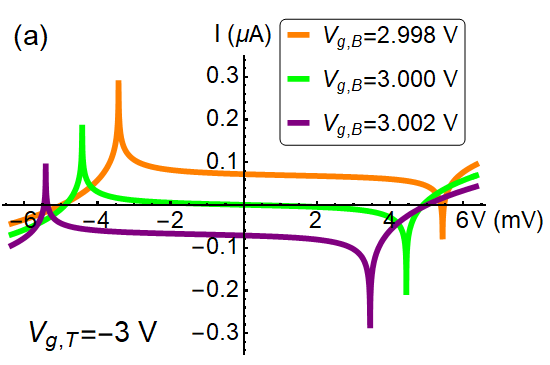}}\qquad
     \subfloat{\includegraphics[scale=0.25]{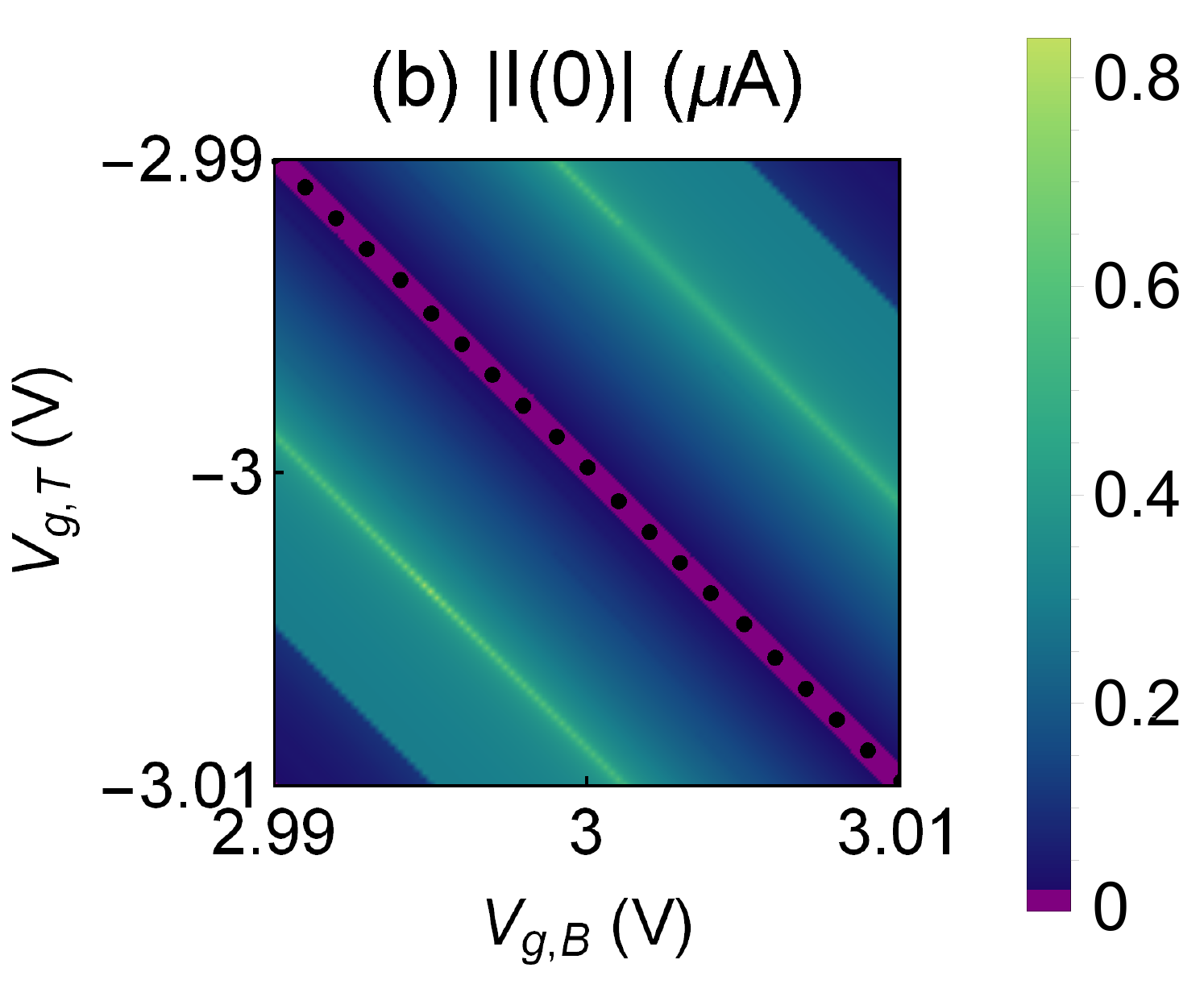}}\qquad
     \subfloat{\includegraphics[scale=0.25]{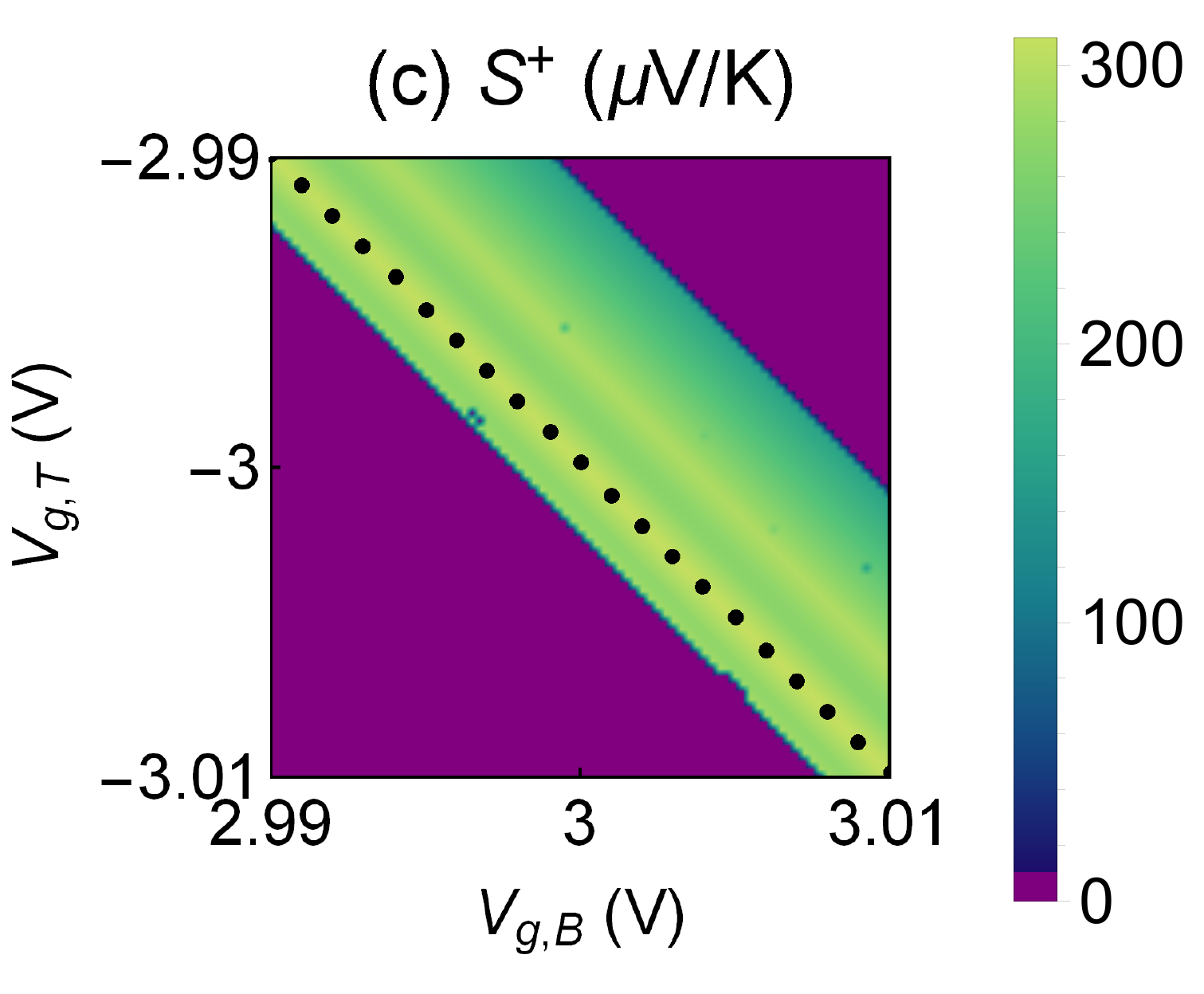}}
     \caption{
     a)IV characteristics for the PH asymmetric state of BLG (orange and violet lines) for $T_L=50\,$K, and $V_{g,T}=-3\,$V. The green line is the PH symmetric case. We notice that these characteristics show a thermoelectric behaviour since in the neighbourhoods of the peaks $P=-IV>0$. b) Current at zero bias (conventional thermocurrent) for $T_L= 50\,$K. Purple areas denote regions where $I(0)\approx 0$, a necessary condition for PH symmetry. c) NSC $S^+$ for $T_L=20\,$K. Within purple regions the junction is not thermoelectric and the NSC cannot be defined. Parameters: $T_R=1~$K, 
     $\varepsilon_b=3.9$, $\varepsilon_t=7.5$, $d_b=100$~nm, $d_t=192$~nm, $d=0.35$~nm. }
     \label{CharNPH}
\end{figure*}
Having treated the PH symmetric case in Sec.~\ref{sec:PHsymmetric}, now we consider the more general case where the BLG is not perfectly gated at the PH symmetric line, i.e., the aforementioned perfect charge neutrality line of the gates.
The loss of PH symmetry on the BLG side, going on with the stated conventions, can be seen as the case where the Fermi energy $E_F^{BLG}\ne 0$. Indeed, according to the self-consistent model of screening, adapted from Refs. \onlinecite{Castro1,McCann1, SciRep}, the electrostatic asymmetry parameter $U$ [and so the bandgap, see Eq.~\eqref{BLGGap}] and the BLG Fermi energy should be computed self-consistently from the value of the electrostatic gating voltages. We state here just the final formula that can be derived from the more general treatments of  Refs. \onlinecite{McCann1,SciRep,Icking2022}.\par
A recent experiment \cite{Icking2022} shows the important role of localized states, in particular for designs with silicon dioxide substrate. These localized states contribute to the charge density of the BLG but not to the screening of electrostatic asymmetry parameter $U$ \footnote{The presence of localized charge states induced by impurities also determines a modification of the gating conditions anyway these contributions can be simply accounted by including appropriate offsets to the gate voltages, so we neglected them~\cite{Icking2022}.}.
The fundamental parameters that will be needed in the IV characteristics are the electrostatic asymmetry parameter $U$ and the Fermi energy $E_F^{BLG}$ that are given by~\footnote{The bandgap self-consistent expression of Eq.~\eqref{eq:Uself} is derived using Eqs.~(65) and (74) of Ref.~\cite{McCann1} in the limit with $\varepsilon_r\sim 1$, $\Lambda\sim 1$ and assuming no mobile charge density when the Fermi energy is in the gap. Correspondingly the Fermi-energy expression Eq.~\eqref{eq:FermiEnergy} coincides with the chemical potential of Eq.~(S13) of the supplementary information of Ref.~\cite{Icking2022} where have been neglected terms involving the inverse of BLG capacitance $C_{BLG}^{-1}$ being $C_{BLG}\gg C_b, C_t$.}
\begin{align}
&U\approx\frac{e\,\bigl(C_b V_{g,B}-C_t V_{g,T}\bigr)}{2C_{BLG}\bigl[1-\ln (|U|/4 t_{\perp})/2\bigr]}\,,
\label{eq:Uself}
\\
&E_F^{BLG}\approx e\frac{\,C_b V_{g,B} + C_t V_{g,T}}{C_b+C_t}\,,
\label{eq:FermiEnergy}
\end{align}
where $C_{BLG}=\varepsilon_0\varepsilon_{r}/d$ is the BLG interlayer capacitance with $d\approx0.35\,$nm the distance of the two graphene layers and $\varepsilon_r\sim 1$~\footnote{This simplified model may not be fully self-consistent when the system is gated well outside the gap, but corrections are expected to slightly renormalize the gap value not affecting physics discussed.}. The solution of the first equation has to be found numerically for every gate configuration $(V_{g,B},V_{g,T})$. The experimental realisation of the device determines the gate capacitances $C_i$ with $i=t,b$ . In order to simplify the gating around the PH symmetric point, it is convenient to consider experimental setups where $C_b\approx C_t$.
In principle, by plugging the these quantities into Eq.~\ref{BLGGap} and Eq.~\ref{Eq:BLGDOS} one can easily determine the BLG's gap and the BLG DoS.\par
In this situation of partial breaking of the PH symmetry, a conventional thermoelectric effect adds to the nonlinear contribution that we have outlined in Sec.~\ref{sec:PHsymmetric}. However, as we discuss below, this conventional contribution does not completely spoil the bipolar thermoelectricity phenomenology. This result shows that the bipolar thermoelectric effect is also stable against slight perturbations of the PH symmetry of the electrodes. The stability is evident by looking at Fig.~\ref{CharNPH}a, where we fix $V_{g,T}=-3\,V$ and slightly change the bottom voltage $V_{g,B}$ moving around the PH symmetry point (green line). In this graph, in fact, we see that the current at the matching peak of the IV characteristics is still  thermoelectric ($IV<0$) even if there is a little departure from the PH symmetry as signalled by a finite linear thermocurrent contribution, i.e.  $I(V=0)\neq 0$. The IV characteristics, in this case, are no longer reciprocal, i.e., $I(V)\neq -I(-V)$, and they do not cross the origin of the axes. Nevertheless, we notice that the system still displays a behaviour connected to the spontaneous breaking of the PH symmetry. Indeed, at $V=0$, the differential conductance $dI/dV$ is negative for both the orange (violet) line cases.
This negative conductance determines an electrical instability. When the system is thermoelectric $IV<0$ this device effectively operates as a battery; when connecting resistive load in parallel, and the junction is able to support two opposite signs for the thermocurrent through the load for the \emph{same} thermal gradient. In fact, if the load resistance is sufficiently high we can find two different solutions for the current circulating in the load (just imposing current conservation in the circuit as in \cite{MarBraGia2}) with opposite sign. This feature demonstrates that the thermoelectric effect is still bipolar, i.e., two opposite signs of the thermocurrent for a given thermal gradient are possible \cite{GaiaNature}. The intrinsic instability at $V\approx 0$ and the bipolar nature of the effect are consequences of the spontaneous breaking of the PH symmetry and would be important in applications such as current-controlled thermoelectrical memories~\cite{GaiaNature,Patent} or even sensors~\cite{paolucci2023}.
\par
It is interesting to investigate how the zero bias thermoelectric current $I(0)$ behaves, fixing a finite temperature gradient, as a function of the voltage of the top and bottom gates, i.e., in the $(V_{g,B},V_{g,T})$ plane. In Fig.~\ref{CharNPH}b we see that this quantity is zero at the PH symmetry line (purple region). The bright yellow lines parallel to that are the lines at which the BLG gap equal the superconductor gap, and so the matching peak is more or less at $V\approx 0$. This interpretation explains why $I(0)$ at these yellow lines is so enhanced. Indeed we notice that the $I(0)$ current can be quite high and even become comparable to the maximal thermocurrent generated by the bipolar thermoelectric effect discussed before. This thermocurrent can be interpreted as that from a conventional thermoelectric effect but presenting also a strong nonlinear enhancement. In other words, we are discussing a thermoelectric behaviour which is linear in the bias $V$ but nonlinear in the thermal gradient.\par

Clearly, When the PH symmetry is broken the IV characteristics is no more anti-symmetric and the bias values $V_S^\pm$, for which $I(V_S^\pm)=0$, are different depending on the sign of $V$, i.e. $|V_S^+|\neq |V_S^-|$. As a consequence, we need to define a NSC which is different depending on the positive/negative voltage branch such that $S^\pm=V_S^\pm/(T_L-T_R)$. 
Considering the NSC $S^+$, for instance, we see that it is zero except for the green-yellow band in Fig. \ref{CharNPH}c. The displayed behavior can be understood as follows.  
In the region of interest $V_{g,B}\approx -V_{g,T}\approx 3$V, the bandgap is approximately constant, reading $E_g\approx U/2\approx 6~$meV, as determined by the self-consistent solution of Eq.~\eqref{eq:Uself}.  
Hence, the overall behaviour of $S^+$  
is mainly determined by the evolution of the Fermi energy [see Eq.~\eqref{eq:FermiEnergy})], which varies significantly once away from the PH symmetry line (black dotted in Fig.~\ref{CharNPH}). This feature makes the green-yellow band asymmetric with respect to the PH symmetry line. When considering $S^{+}$, we compute the zero-current state with $V>0$. For $|V_{g,B}|<|V_{g,T}|$, the current-voltage characteristic shifts towards higher current values (cf. \ref{CharNPH}a), so making the thermoelectric peak not intersecting the $I=0$ axes above a threshold value of $E_F^{BLG}$. 
Differently, for $|V_{g,B}|>|V_{g,T}|$, the characteristic is shifted in the opposite direction, and the intersection persists until the matching peak condition is at $V=0$.  
Since the value of the NSC mainly depends on $E_g$, as discussed above, we notice that within 
the green and yellow region, $S_+$ depends only weakly on the gates (evolution determined by variation of the Fermi energy). The maximum NSC reaches $300\,\mu \rm{V/K}$ for $T_L=20\,$K, in agreement with Fig.~\ref{WP}d.
Finally, we observe that these figures show that small deviations from the PH symmetry would not completely spoil the bipolar nonlinear thermoelectric effect, although one clearly loses the full antisymmetry of the IV characteristics. This fact confirms the robustness of the discussed bipolar thermoelectric effect, since it results to be stable against small deviations from ideal gating conditions.
\par

\section{Conclusions}

In this article, we analysed the nonlinear thermoelectricity of a BLG-Superconductor tunnelling junction. We demonstrated that by external gating it is possible to tune the BLG gap and the thermoelectric effect by only electrical means, achieving very promising values for the figures of merit, e.g., NSC up to $1\,$mV/K at temperature of the BLG of $T_L=10\,$K. For such temperature gradients, we predicted a thermoelectric generation of power of hundreds of pW for $0.1\,\mu{\rm m}^2$ of active  surface, which corresponds to a surface power density of roughly $1 \,{\rm nW}/\mu {\rm m}^2$ with a maximal absolute thermodynamic efficiency up to $40\%$ for the electronic degrees of freedom.
Finally, we investigated how the different figures of merit depend on the BLG electronic temperature and on the BLG gap in a realistic range of variations of external parameters.
The stated bipolar thermoelectric phenomenology could survive even to a partial breaking of the PH symmetry, as might be potentially induced by a non-ideal external electrostatic gating of the BLG electrode. We discussed, for this more general case as well, the thermoelectric properties of the device, showing indeed that there is an interplay between the nonlinear bipolar thermoelectricity induced by spontaneous breaking of the PH symmetry and the conventional unipolar thermoelectricity as determined by a small PH breaking term. However, we confirmed that this latter contribution can coexist with the former without being detrimental for the bipolar thermoelectricity. We state one more time that the phenomenology just outlined could lead to a wide versatility of this sort of devices due to the tunability provided by the external gating mechanism.\par

Finally, we hope that the sizable thermoelectric performances of this hybrid device architecture, combined with its electrical flexibility, will trigger the future experimental exploration of bipolar thermoelectricity at even higher temperatures. Indeed, the tunability of the gap makes the device adaptable to respond to unwanted changes in the temperature of the hot lead and, at the same time, can be exploited to switch on and off its thermoelectric behaviour.
We have seen that hybrid superconducting-graphene devices could play an important role in calorimetry and fast bolometry \cite{Xia2009, Mueller2010,Lee2020,Vischi}. We could envision similar applications for our device \cite{paolucci2023}. Furthermore, one could also exploit our system in thermoelectric nanodevices similar to those designed in different platforms \cite{Varpula2017,Heikkila2018,Geng2020,Varpula2021} carrying the advantage of  higher operating temperatures. To conclude, we hope that the peculiar features of the bipolar thermoelectric effect here reported will spur further applications in the field of quantum technologies \cite{Patent}.

\begin{acknowledgments}

The authors wish to thank for useful discussion Dr. A. Tomadin, Dr. T. Novotny, Dr. K. Michaeli, and Prof. F. Strocchi. F.G. and A.B. acknowledge the EU's Horizon 2020 research and innovation program under Grant Agreement No. 800923 (SUPERTED) and No. 964398 (SUPERGATE) for partial financial support.
S.R. acknowledges the financial support of the Italian Ministry of University and Research (PRIN project QUANTUM2D).
A.B. acknowledges the Royal Society through the International Exchanges between the UK and Italy (Grants No. IEC R2 192166 and IEC R2 212041)
\end{acknowledgments}
\nocite{*}
\bibliography{references}

\providecommand{\noopsort}[1]{}\providecommand{\singleletter}[1]{#1}%
\begin{thebibliography}{83}%
\makeatletter
\providecommand \@ifxundefined [1]{%
 \@ifx{#1\undefined}
}%
\providecommand \@ifnum [1]{%
 \ifnum #1\expandafter \@firstoftwo
 \else \expandafter \@secondoftwo
 \fi
}%
\providecommand \@ifx [1]{%
 \ifx #1\expandafter \@firstoftwo
 \else \expandafter \@secondoftwo
 \fi
}%
\providecommand \natexlab [1]{#1}%
\providecommand \enquote  [1]{``#1''}%
\providecommand \bibnamefont  [1]{#1}%
\providecommand \bibfnamefont [1]{#1}%
\providecommand \citenamefont [1]{#1}%
\providecommand \href@noop [0]{\@secondoftwo}%
\providecommand \href [0]{\begingroup \@sanitize@url \@href}%
\providecommand \@href[1]{\@@startlink{#1}\@@href}%
\providecommand \@@href[1]{\endgroup#1\@@endlink}%
\providecommand \@sanitize@url [0]{\catcode `\\12\catcode `\$12\catcode
  `\&12\catcode `\#12\catcode `\^12\catcode `\_12\catcode `\%12\relax}%
\providecommand \@@startlink[1]{}%
\providecommand \@@endlink[0]{}%
\providecommand \url  [0]{\begingroup\@sanitize@url \@url }%
\providecommand \@url [1]{\endgroup\@href {#1}{\urlprefix }}%
\providecommand \urlprefix  [0]{URL }%
\providecommand \Eprint [0]{\href }%
\providecommand \doibase [0]{https://doi.org/}%
\providecommand \selectlanguage [0]{\@gobble}%
\providecommand \bibinfo  [0]{\@secondoftwo}%
\providecommand \bibfield  [0]{\@secondoftwo}%
\providecommand \translation [1]{[#1]}%
\providecommand \BibitemOpen [0]{}%
\providecommand \bibitemStop [0]{}%
\providecommand \bibitemNoStop [0]{.\EOS\space}%
\providecommand \EOS [0]{\spacefactor3000\relax}%
\providecommand \BibitemShut  [1]{\csname bibitem#1\endcsname}%
\let\auto@bib@innerbib\@empty
\bibitem [{\citenamefont {Tritt}(2004)}]{Tritt}%
  \BibitemOpen
  \bibfield  {author} {\bibinfo {author} {\bibfnamefont {T.~M.}\ \bibnamefont
  {Tritt}},\ }\href@noop {} {\emph {\bibinfo {title} {Thermal Conductivity:
  theory, properties, and applications}}}\ (\bibinfo  {publisher} {Kluwer
  Academic/Plenum Publishers, New York},\ \bibinfo {year} {2004})\BibitemShut
  {NoStop}%
\bibitem [{\citenamefont {Tritt}\ and\ \citenamefont
  {Subramanian}(2006)}]{TrittSub}%
  \BibitemOpen
  \bibfield  {author} {\bibinfo {author} {\bibfnamefont {T.~M.}\ \bibnamefont
  {Tritt}}\ and\ \bibinfo {author} {\bibfnamefont {M.~A.}\ \bibnamefont
  {Subramanian}},\ }\bibfield  {title} {\bibinfo {title} {Thermoelectric
  materials, phenomena, and applications: a bird's eye view},\ }\href@noop {}
  {\bibfield  {journal} {\bibinfo  {journal} {MRS Bulletin}\ }\textbf {\bibinfo
  {volume} {31}},\ \bibinfo {pages} {188} (\bibinfo {year} {2006})}\BibitemShut
  {NoStop}%
\bibitem [{\citenamefont {Macià-Barber}(2015)}]{Barber}%
  \BibitemOpen
  \bibfield  {author} {\bibinfo {author} {\bibfnamefont {E.}~\bibnamefont
  {Macià-Barber}},\ }\href@noop {} {\emph {\bibinfo {title} {Thermoelectric
  Materials: Advances and Applications}}}\ (\bibinfo  {publisher} {Pan Stanford
  Publishing, Taylor and Francis Group},\ \bibinfo {year} {2015})\BibitemShut
  {NoStop}%
\bibitem [{\citenamefont {Grosso}\ and\ \citenamefont
  {Parravicini}(2014)}]{Grosso}%
  \BibitemOpen
  \bibfield  {author} {\bibinfo {author} {\bibfnamefont {G.}~\bibnamefont
  {Grosso}}\ and\ \bibinfo {author} {\bibfnamefont {G.~P.}\ \bibnamefont
  {Parravicini}},\ }\href@noop {} {\emph {\bibinfo {title} {Solid State
  Physics}}},\ \bibinfo {edition} {2nd}\ ed.\ (\bibinfo  {publisher} {Academic
  Press, Elsevier Ltd.},\ \bibinfo {year} {2014})\BibitemShut {NoStop}%
\bibitem [{\citenamefont {Benenti}\ \emph {et~al.}(2017)\citenamefont
  {Benenti}, \citenamefont {Casati}, \citenamefont {Saito},\ and\ \citenamefont
  {Whitney}}]{Benenti}%
  \BibitemOpen
  \bibfield  {author} {\bibinfo {author} {\bibfnamefont {G.}~\bibnamefont
  {Benenti}}, \bibinfo {author} {\bibfnamefont {G.}~\bibnamefont {Casati}},
  \bibinfo {author} {\bibfnamefont {K.}~\bibnamefont {Saito}},\ and\ \bibinfo
  {author} {\bibfnamefont {R.}~\bibnamefont {Whitney}},\ }\bibfield  {title}
  {\bibinfo {title} {Fundamental aspects of steady-state conversion of heat to
  work at the nanoscale},\ }\href@noop {} {\bibfield  {journal} {\bibinfo
  {journal} {Phys. Rep.}\ }\textbf {\bibinfo {volume} {694}},\ \bibinfo {pages}
  {1} (\bibinfo {year} {2017})}\BibitemShut {NoStop}%
\bibitem [{\citenamefont {Ginzburg}(1991)}]{Ginzburg}%
  \BibitemOpen
  \bibfield  {author} {\bibinfo {author} {\bibfnamefont {V.}~\bibnamefont
  {Ginzburg}},\ }\bibfield  {title} {\bibinfo {title} {Thermoelectric effects
  in the superconducting state},\ }\href@noop {} {\bibfield  {journal}
  {\bibinfo  {journal} {Sov. Phys. Usp.}\ }\textbf {\bibinfo {volume} {34}},\
  \bibinfo {pages} {101} (\bibinfo {year} {1991})}\BibitemShut {NoStop}%
\bibitem [{\citenamefont {Smith}\ \emph {et~al.}(1980)\citenamefont {Smith},
  \citenamefont {Tinkham},\ and\ \citenamefont {Skocpol}}]{SmithTinkham}%
  \BibitemOpen
  \bibfield  {author} {\bibinfo {author} {\bibfnamefont {A.~D.}\ \bibnamefont
  {Smith}}, \bibinfo {author} {\bibfnamefont {M.}~\bibnamefont {Tinkham}},\
  and\ \bibinfo {author} {\bibfnamefont {W.~J.}\ \bibnamefont {Skocpol}},\
  }\bibfield  {title} {\bibinfo {title} {New thermoelectric effect in tunnel
  junctions},\ }\href@noop {} {\bibfield  {journal} {\bibinfo  {journal} {Phys.
  Rev. B}\ }\textbf {\bibinfo {volume} {22}},\ \bibinfo {pages} {4346}
  (\bibinfo {year} {1980})}\BibitemShut {NoStop}%
\bibitem [{\citenamefont {Machon}\ \emph {et~al.}(2013)\citenamefont {Machon},
  \citenamefont {Eschrig},\ and\ \citenamefont {Belzig}}]{Machon13}%
  \BibitemOpen
  \bibfield  {author} {\bibinfo {author} {\bibfnamefont {P.}~\bibnamefont
  {Machon}}, \bibinfo {author} {\bibfnamefont {M.}~\bibnamefont {Eschrig}},\
  and\ \bibinfo {author} {\bibfnamefont {W.}~\bibnamefont {Belzig}},\
  }\bibfield  {title} {\bibinfo {title} {Nonlocal thermoelectric effects and
  nonlocal onsager relations in a three-terminal proximity-coupled
  superconductor-ferromagnet device},\ }\href@noop {} {\bibfield  {journal}
  {\bibinfo  {journal} {Phys. Rev. Lett.}\ }\textbf {\bibinfo {volume} {110}},\
  \bibinfo {pages} {047002} (\bibinfo {year} {2013})}\BibitemShut {NoStop}%
\bibitem [{\citenamefont {Ozaeta}\ \emph {et~al.}(2014)\citenamefont {Ozaeta},
  \citenamefont {Virtanen}, \citenamefont {Bergeret},\ and\ \citenamefont
  {Heikkil\"a}}]{Ozaeta14}%
  \BibitemOpen
  \bibfield  {author} {\bibinfo {author} {\bibfnamefont {A.}~\bibnamefont
  {Ozaeta}}, \bibinfo {author} {\bibfnamefont {P.}~\bibnamefont {Virtanen}},
  \bibinfo {author} {\bibfnamefont {F.~S.}\ \bibnamefont {Bergeret}},\ and\
  \bibinfo {author} {\bibfnamefont {T.~T.}\ \bibnamefont {Heikkil\"a}},\
  }\bibfield  {title} {\bibinfo {title} {Predicted very large thermoelectric
  effect in ferromagnet-superconductor junctions in the presence of a
  spin-splitting magnetic field},\ }\href@noop {} {\bibfield  {journal}
  {\bibinfo  {journal} {Phys. Rev. Lett.}\ }\textbf {\bibinfo {volume} {112}},\
  \bibinfo {pages} {057001} (\bibinfo {year} {2014})}\BibitemShut {NoStop}%
\bibitem [{\citenamefont {S\'anchez}\ \emph {et~al.}(2018)\citenamefont
  {S\'anchez}, \citenamefont {Burset},\ and\ \citenamefont
  {Yeyati}}]{SanchezML2}%
  \BibitemOpen
  \bibfield  {author} {\bibinfo {author} {\bibfnamefont {R.}~\bibnamefont
  {S\'anchez}}, \bibinfo {author} {\bibfnamefont {P.}~\bibnamefont {Burset}},\
  and\ \bibinfo {author} {\bibfnamefont {A.~L.}\ \bibnamefont {Yeyati}},\
  }\bibfield  {title} {\bibinfo {title} {Cooling by cooper pair splitting},\
  }\href@noop {} {\bibfield  {journal} {\bibinfo  {journal} {Phys. Rev. B}\
  }\textbf {\bibinfo {volume} {98}},\ \bibinfo {pages} {241414} (\bibinfo
  {year} {2018})}\BibitemShut {NoStop}%
\bibitem [{\citenamefont {Tabatabaei}\ \emph {et~al.}(2022)\citenamefont
  {Tabatabaei}, \citenamefont {S\'anchez}, \citenamefont {Yeyati},\ and\
  \citenamefont {S\'anchez}}]{SanchezML1}%
  \BibitemOpen
  \bibfield  {author} {\bibinfo {author} {\bibfnamefont {S.~M.}\ \bibnamefont
  {Tabatabaei}}, \bibinfo {author} {\bibfnamefont {D.}~\bibnamefont
  {S\'anchez}}, \bibinfo {author} {\bibfnamefont {A.~L.}\ \bibnamefont
  {Yeyati}},\ and\ \bibinfo {author} {\bibfnamefont {R.}~\bibnamefont
  {S\'anchez}},\ }\bibfield  {title} {\bibinfo {title} {Nonlocal quantum heat
  engines made of hybrid superconducting devices},\ }\href@noop {} {\bibfield
  {journal} {\bibinfo  {journal} {Phys. Rev. B}\ }\textbf {\bibinfo {volume}
  {106}},\ \bibinfo {pages} {115419} (\bibinfo {year} {2022})}\BibitemShut
  {NoStop}%
\bibitem [{\citenamefont {Pershoguba}\ and\ \citenamefont
  {Glazman}(2019)}]{Pershoguba19}%
  \BibitemOpen
  \bibfield  {author} {\bibinfo {author} {\bibfnamefont {S.~S.}\ \bibnamefont
  {Pershoguba}}\ and\ \bibinfo {author} {\bibfnamefont {L.~I.}\ \bibnamefont
  {Glazman}},\ }\bibfield  {title} {\bibinfo {title} {Thermopower and thermal
  conductance of a superconducting quantum point contact},\ }\href@noop {}
  {\bibfield  {journal} {\bibinfo  {journal} {Phys. Rev. B}\ }\textbf {\bibinfo
  {volume} {99}},\ \bibinfo {pages} {134514} (\bibinfo {year}
  {2019})}\BibitemShut {NoStop}%
\bibitem [{\citenamefont {Virtanen}\ and\ \citenamefont
  {Heikkil\"a}(2004)}]{Virtanen04}%
  \BibitemOpen
  \bibfield  {author} {\bibinfo {author} {\bibfnamefont {P.}~\bibnamefont
  {Virtanen}}\ and\ \bibinfo {author} {\bibfnamefont {T.}~\bibnamefont
  {Heikkil\"a}},\ }\bibfield  {title} {\bibinfo {title} {Thermopower in andreev
  interferometers},\ }\href@noop {} {\bibfield  {journal} {\bibinfo  {journal}
  {J. Low Temp. Phys.}\ }\textbf {\bibinfo {volume} {136}},\ \bibinfo {pages}
  {625} (\bibinfo {year} {2004})}\BibitemShut {NoStop}%
\bibitem [{\citenamefont {Jacquod}\ and\ \citenamefont
  {Whitney}(2010)}]{Jacquod10}%
  \BibitemOpen
  \bibfield  {author} {\bibinfo {author} {\bibfnamefont {P.}~\bibnamefont
  {Jacquod}}\ and\ \bibinfo {author} {\bibfnamefont {R.}~\bibnamefont
  {Whitney}},\ }\bibfield  {title} {\bibinfo {title} {Coherent thermoelectric
  effects in mesoscopic andreev interferometers},\ }\href@noop {} {\bibfield
  {journal} {\bibinfo  {journal} {Europhys. Lett.}\ }\textbf {\bibinfo {volume}
  {91}},\ \bibinfo {pages} {67009} (\bibinfo {year} {2010})}\BibitemShut
  {NoStop}%
\bibitem [{\citenamefont {Kalenkov}\ and\ \citenamefont
  {Zaikin}(2017)}]{Kalenkov17}%
  \BibitemOpen
  \bibfield  {author} {\bibinfo {author} {\bibfnamefont {M.~S.}\ \bibnamefont
  {Kalenkov}}\ and\ \bibinfo {author} {\bibfnamefont {A.~D.}\ \bibnamefont
  {Zaikin}},\ }\bibfield  {title} {\bibinfo {title} {Large thermoelectric
  effect in ballistic andreev interferometers},\ }\href@noop {} {\bibfield
  {journal} {\bibinfo  {journal} {Phys. Rev. B}\ }\textbf {\bibinfo {volume}
  {95}},\ \bibinfo {pages} {024518} (\bibinfo {year} {2017})}\BibitemShut
  {NoStop}%
\bibitem [{\citenamefont {Hussein}\ \emph {et~al.}(2019)\citenamefont
  {Hussein}, \citenamefont {Governale}, \citenamefont {Kohler}, \citenamefont
  {Belzig}, \citenamefont {Giazotto},\ and\ \citenamefont
  {Braggio}}]{Hussein19}%
  \BibitemOpen
  \bibfield  {author} {\bibinfo {author} {\bibfnamefont {R.}~\bibnamefont
  {Hussein}}, \bibinfo {author} {\bibfnamefont {M.}~\bibnamefont {Governale}},
  \bibinfo {author} {\bibfnamefont {S.}~\bibnamefont {Kohler}}, \bibinfo
  {author} {\bibfnamefont {W.}~\bibnamefont {Belzig}}, \bibinfo {author}
  {\bibfnamefont {F.}~\bibnamefont {Giazotto}},\ and\ \bibinfo {author}
  {\bibfnamefont {A.}~\bibnamefont {Braggio}},\ }\bibfield  {title} {\bibinfo
  {title} {Nonlocal thermoelectricity in a cooper-pair splitter},\ }\href@noop
  {} {\bibfield  {journal} {\bibinfo  {journal} {Phys. Rev. B}\ }\textbf
  {\bibinfo {volume} {99}},\ \bibinfo {pages} {075429} (\bibinfo {year}
  {2019})}\BibitemShut {NoStop}%
\bibitem [{\citenamefont {Kirsanov}\ \emph {et~al.}(2019)\citenamefont
  {Kirsanov}, \citenamefont {Tan}, \citenamefont {Golubev}, \citenamefont
  {Hakonen},\ and\ \citenamefont {Lesovik}}]{Kirsanov19}%
  \BibitemOpen
  \bibfield  {author} {\bibinfo {author} {\bibfnamefont {N.~S.}\ \bibnamefont
  {Kirsanov}}, \bibinfo {author} {\bibfnamefont {Z.~B.}\ \bibnamefont {Tan}},
  \bibinfo {author} {\bibfnamefont {D.~S.}\ \bibnamefont {Golubev}}, \bibinfo
  {author} {\bibfnamefont {P.~J.}\ \bibnamefont {Hakonen}},\ and\ \bibinfo
  {author} {\bibfnamefont {G.~B.}\ \bibnamefont {Lesovik}},\ }\bibfield
  {title} {\bibinfo {title} {Heat switch and thermoelectric effects based on
  cooper-pair splitting and elastic cotunneling},\ }\href@noop {} {\bibfield
  {journal} {\bibinfo  {journal} {Phys. Rev. B}\ }\textbf {\bibinfo {volume}
  {99}},\ \bibinfo {pages} {115127} (\bibinfo {year} {2019})}\BibitemShut
  {NoStop}%
\bibitem [{\citenamefont {Ratnakar}\ and\ \citenamefont {Das}(2021)}]{Das21}%
  \BibitemOpen
  \bibfield  {author} {\bibinfo {author} {\bibfnamefont {A.}~\bibnamefont
  {Ratnakar}}\ and\ \bibinfo {author} {\bibfnamefont {S.}~\bibnamefont {Das}},\
  }\bibfield  {title} {\bibinfo {title} {Enhancement in tunneling density of
  states in a luttinger liquid: role of nonlocal interaction},\ }\href@noop {}
  {\bibfield  {journal} {\bibinfo  {journal} {Phys. Rev. B}\ }\textbf {\bibinfo
  {volume} {104}},\ \bibinfo {pages} {045402} (\bibinfo {year}
  {2021})}\BibitemShut {NoStop}%
\bibitem [{\citenamefont {Eom}\ \emph {et~al.}(1998)\citenamefont {Eom},
  \citenamefont {Chien},\ and\ \citenamefont {Chandrasekhar}}]{Eom98}%
  \BibitemOpen
  \bibfield  {author} {\bibinfo {author} {\bibfnamefont {J.}~\bibnamefont
  {Eom}}, \bibinfo {author} {\bibfnamefont {C.-J.}\ \bibnamefont {Chien}},\
  and\ \bibinfo {author} {\bibfnamefont {V.}~\bibnamefont {Chandrasekhar}},\
  }\bibfield  {title} {\bibinfo {title} {Phase dependent thermopower in andreev
  interferometers},\ }\href@noop {} {\bibfield  {journal} {\bibinfo  {journal}
  {Phys. Rev. Lett.}\ }\textbf {\bibinfo {volume} {81}},\ \bibinfo {pages}
  {437} (\bibinfo {year} {1998})}\BibitemShut {NoStop}%
\bibitem [{\citenamefont {Jiang}\ and\ \citenamefont
  {Chandrasekhar}(2005)}]{Jiang05}%
  \BibitemOpen
  \bibfield  {author} {\bibinfo {author} {\bibfnamefont {Z.}~\bibnamefont
  {Jiang}}\ and\ \bibinfo {author} {\bibfnamefont {V.}~\bibnamefont
  {Chandrasekhar}},\ }\bibfield  {title} {\bibinfo {title} {Quantitative
  measurements of the thermal resistance of andreev interferometers},\
  }\href@noop {} {\bibfield  {journal} {\bibinfo  {journal} {Phys. Rev. B}\
  }\textbf {\bibinfo {volume} {72}},\ \bibinfo {pages} {020502} (\bibinfo
  {year} {2005})}\BibitemShut {NoStop}%
\bibitem [{\citenamefont {Tan}\ \emph {et~al.}(2021)\citenamefont {Tan},
  \citenamefont {Kirsanov}, \citenamefont {Galda}, \citenamefont {Vinokur},
  \citenamefont {Haque}, \citenamefont {Savin}, \citenamefont {Golubev},
  \citenamefont {Lesovik},\ and\ \citenamefont {Hakonen}}]{Tan21}%
  \BibitemOpen
  \bibfield  {author} {\bibinfo {author} {\bibfnamefont {Z.}~\bibnamefont
  {Tan}}, \bibinfo {author} {\bibfnamefont {A.~L.~N.}\ \bibnamefont
  {Kirsanov}}, \bibinfo {author} {\bibfnamefont {A.}~\bibnamefont {Galda}},
  \bibinfo {author} {\bibfnamefont {V.}~\bibnamefont {Vinokur}}, \bibinfo
  {author} {\bibfnamefont {M.}~\bibnamefont {Haque}}, \bibinfo {author}
  {\bibfnamefont {A.}~\bibnamefont {Savin}}, \bibinfo {author} {\bibfnamefont
  {D.}~\bibnamefont {Golubev}}, \bibinfo {author} {\bibfnamefont
  {G.}~\bibnamefont {Lesovik}},\ and\ \bibinfo {author} {\bibfnamefont
  {P.}~\bibnamefont {Hakonen}},\ }\bibfield  {title} {\bibinfo {title}
  {Thermoelectric current in a graphene cooper pair splitter.},\ }\href@noop {}
  {\bibfield  {journal} {\bibinfo  {journal} {Nat. Commun.}\ }\textbf {\bibinfo
  {volume} {12}},\ \bibinfo {pages} {138} (\bibinfo {year} {2021})}\BibitemShut
  {NoStop}%
\bibitem [{\citenamefont {Aronov}\ and\ \citenamefont {Spivak}(1975)}]{Spivak}%
  \BibitemOpen
  \bibfield  {author} {\bibinfo {author} {\bibfnamefont {A.~G.}\ \bibnamefont
  {Aronov}}\ and\ \bibinfo {author} {\bibfnamefont {B.~Z.}\ \bibnamefont
  {Spivak}},\ }\bibfield  {title} {\bibinfo {title} {Photoeffect in a josephson
  junction},\ }\href@noop {} {\bibfield  {journal} {\bibinfo  {journal} {JETP
  Lett.}\ }\textbf {\bibinfo {volume} {22}},\ \bibinfo {pages} {101} (\bibinfo
  {year} {1975})}\BibitemShut {NoStop}%
\bibitem [{\citenamefont {Gershenzon}\ and\ \citenamefont
  {Falei}(1986)}]{Gershenzon1}%
  \BibitemOpen
  \bibfield  {author} {\bibinfo {author} {\bibfnamefont {M.~E.}\ \bibnamefont
  {Gershenzon}}\ and\ \bibinfo {author} {\bibfnamefont {M.~I.}\ \bibnamefont
  {Falei}},\ }\bibfield  {title} {\bibinfo {title} {Absolute negative
  resistance of a tunnel contact between superconductors with a nonequilibrium
  quasiparticle distribution function},\ }\href@noop {} {\bibfield  {journal}
  {\bibinfo  {journal} {JETP Lett.}\ }\textbf {\bibinfo {volume} {44}},\
  \bibinfo {pages} {682} (\bibinfo {year} {1986})}\BibitemShut {NoStop}%
\bibitem [{\citenamefont {Gershenzon}\ and\ \citenamefont
  {Falei}(1988)}]{Gershenzon2}%
  \BibitemOpen
  \bibfield  {author} {\bibinfo {author} {\bibfnamefont {M.~E.}\ \bibnamefont
  {Gershenzon}}\ and\ \bibinfo {author} {\bibfnamefont {M.~I.}\ \bibnamefont
  {Falei}},\ }\bibfield  {title} {\bibinfo {title} {Absolute negative
  resistance in tunnel junctions of nonequilibrium superconductors},\
  }\href@noop {} {\bibfield  {journal} {\bibinfo  {journal} {Sov. Phys. JETP}\
  }\textbf {\bibinfo {volume} {67}},\ \bibinfo {pages} {389} (\bibinfo {year}
  {1988})}\BibitemShut {NoStop}%
\bibitem [{\citenamefont {Gijsbertsen}\ and\ \citenamefont
  {Flokstra}(1996)}]{Flokstra}%
  \BibitemOpen
  \bibfield  {author} {\bibinfo {author} {\bibfnamefont {J.~G.}\ \bibnamefont
  {Gijsbertsen}}\ and\ \bibinfo {author} {\bibfnamefont {J.}~\bibnamefont
  {Flokstra}},\ }\bibfield  {title} {\bibinfo {title} {Quasiparticle
  injection‐detection experiments in niobium},\ }\href@noop {} {\bibfield
  {journal} {\bibinfo  {journal} {J. Appl. Phys.}\ }\textbf {\bibinfo {volume}
  {80}},\ \bibinfo {pages} {3923} (\bibinfo {year} {1996})}\BibitemShut
  {NoStop}%
\bibitem [{\citenamefont {Marchegiani}\ \emph
  {et~al.}(2020{\natexlab{a}})\citenamefont {Marchegiani}, \citenamefont
  {Braggio},\ and\ \citenamefont {Giazotto}}]{MarBraGia1}%
  \BibitemOpen
  \bibfield  {author} {\bibinfo {author} {\bibfnamefont {G.}~\bibnamefont
  {Marchegiani}}, \bibinfo {author} {\bibfnamefont {A.}~\bibnamefont
  {Braggio}},\ and\ \bibinfo {author} {\bibfnamefont {F.}~\bibnamefont
  {Giazotto}},\ }\bibfield  {title} {\bibinfo {title} {Nonlinear
  thermoelectricity with electron-hole symmetric systems},\ }\href@noop {}
  {\bibfield  {journal} {\bibinfo  {journal} {Phys. Rev. Lett.}\ }\textbf
  {\bibinfo {volume} {124}},\ \bibinfo {pages} {106801} (\bibinfo {year}
  {2020}{\natexlab{a}})}\BibitemShut {NoStop}%
\bibitem [{\citenamefont {Marchegiani}\ \emph
  {et~al.}(2020{\natexlab{b}})\citenamefont {Marchegiani}, \citenamefont
  {Braggio},\ and\ \citenamefont {Giazotto}}]{MarBraGia2}%
  \BibitemOpen
  \bibfield  {author} {\bibinfo {author} {\bibfnamefont {G.}~\bibnamefont
  {Marchegiani}}, \bibinfo {author} {\bibfnamefont {A.}~\bibnamefont
  {Braggio}},\ and\ \bibinfo {author} {\bibfnamefont {F.}~\bibnamefont
  {Giazotto}},\ }\bibfield  {title} {\bibinfo {title} {Superconducting
  nonlinear thermoelectric heat engine},\ }\href@noop {} {\bibfield  {journal}
  {\bibinfo  {journal} {Phys. Rev. B}\ }\textbf {\bibinfo {volume} {101}},\
  \bibinfo {pages} {214509} (\bibinfo {year} {2020}{\natexlab{b}})}\BibitemShut
  {NoStop}%
\bibitem [{\citenamefont {Marchegiani}\ \emph
  {et~al.}(2020{\natexlab{c}})\citenamefont {Marchegiani}, \citenamefont
  {Braggio},\ and\ \citenamefont {Giazotto}}]{MarBraGia3}%
  \BibitemOpen
  \bibfield  {author} {\bibinfo {author} {\bibfnamefont {G.}~\bibnamefont
  {Marchegiani}}, \bibinfo {author} {\bibfnamefont {A.}~\bibnamefont
  {Braggio}},\ and\ \bibinfo {author} {\bibfnamefont {F.}~\bibnamefont
  {Giazotto}},\ }\bibfield  {title} {\bibinfo {title} {Phase-tunable
  thermoelectricity in a josephson junction},\ }\href@noop {} {\bibfield
  {journal} {\bibinfo  {journal} {Phys. Rev. Research}\ }\textbf {\bibinfo
  {volume} {2}},\ \bibinfo {pages} {043091} (\bibinfo {year}
  {2020}{\natexlab{c}})}\BibitemShut {NoStop}%
\bibitem [{\citenamefont {Marchegiani}\ \emph
  {et~al.}(2020{\natexlab{d}})\citenamefont {Marchegiani}, \citenamefont
  {Braggio},\ and\ \citenamefont {Giazotto}}]{MarBraGia4}%
  \BibitemOpen
  \bibfield  {author} {\bibinfo {author} {\bibfnamefont {G.}~\bibnamefont
  {Marchegiani}}, \bibinfo {author} {\bibfnamefont {A.}~\bibnamefont
  {Braggio}},\ and\ \bibinfo {author} {\bibfnamefont {F.}~\bibnamefont
  {Giazotto}},\ }\bibfield  {title} {\bibinfo {title} {Noise effects in the
  nonlinear thermoelectricity of a josephson junction},\ }\href@noop {}
  {\bibfield  {journal} {\bibinfo  {journal} {Appl. Phys. Lett.}\ }\textbf
  {\bibinfo {volume} {117}},\ \bibinfo {pages} {212601} (\bibinfo {year}
  {2020}{\natexlab{d}})}\BibitemShut {NoStop}%
\bibitem [{\citenamefont {Germanese}\ \emph {et~al.}(2022)\citenamefont
  {Germanese}, \citenamefont {Paolucci}, \citenamefont {Marchegiani},
  \citenamefont {Braggio},\ and\ \citenamefont {Giazotto}}]{GaiaNature}%
  \BibitemOpen
  \bibfield  {author} {\bibinfo {author} {\bibfnamefont {G.}~\bibnamefont
  {Germanese}}, \bibinfo {author} {\bibfnamefont {F.}~\bibnamefont {Paolucci}},
  \bibinfo {author} {\bibfnamefont {G.}~\bibnamefont {Marchegiani}}, \bibinfo
  {author} {\bibfnamefont {A.}~\bibnamefont {Braggio}},\ and\ \bibinfo {author}
  {\bibfnamefont {F.}~\bibnamefont {Giazotto}},\ }\bibfield  {title} {\bibinfo
  {title} {Bipolar thermoelectric josephson engine},\ }\href
  {https://doi.org/10.1038/s41565-022-01208-y} {\bibfield  {journal} {\bibinfo
  {journal} {Nat. Nanotechnol.}\ }\textbf {\bibinfo {volume} {17}},\ \bibinfo
  {pages} {1084} (\bibinfo {year} {2022})}\BibitemShut {NoStop}%
\bibitem [{\citenamefont {Germanese}\ \emph {et~al.}(2023)\citenamefont
  {Germanese}, \citenamefont {Paolucci}, \citenamefont {Marchegiani},
  \citenamefont {Braggio},\ and\ \citenamefont {Giazotto}}]{GaiaPRApp}%
  \BibitemOpen
  \bibfield  {author} {\bibinfo {author} {\bibfnamefont {G.}~\bibnamefont
  {Germanese}}, \bibinfo {author} {\bibfnamefont {F.}~\bibnamefont {Paolucci}},
  \bibinfo {author} {\bibfnamefont {G.}~\bibnamefont {Marchegiani}}, \bibinfo
  {author} {\bibfnamefont {A.}~\bibnamefont {Braggio}},\ and\ \bibinfo {author}
  {\bibfnamefont {F.}~\bibnamefont {Giazotto}},\ }\bibfield  {title} {\bibinfo
  {title} {Phase control of bipolar thermoelectricity in josephson tunnel
  junctions},\ }\href {https://doi.org/10.1103/PhysRevApplied.19.014074}
  {\bibfield  {journal} {\bibinfo  {journal} {Phys. Rev. Appl.}\ }\textbf
  {\bibinfo {volume} {19}},\ \bibinfo {pages} {014074} (\bibinfo {year}
  {2023})}\BibitemShut {NoStop}%
\bibitem [{\citenamefont {Giazotto}\ \emph {et~al.}(2021)\citenamefont
  {Giazotto}, \citenamefont {Paolucci}, \citenamefont {Braggio}, \citenamefont
  {Marchegiani},\ and\ \citenamefont {Germanese}}]{Patent}%
  \BibitemOpen
  \bibfield  {author} {\bibinfo {author} {\bibfnamefont {F.}~\bibnamefont
  {Giazotto}}, \bibinfo {author} {\bibfnamefont {F.}~\bibnamefont {Paolucci}},
  \bibinfo {author} {\bibfnamefont {A.}~\bibnamefont {Braggio}}, \bibinfo
  {author} {\bibfnamefont {G.}~\bibnamefont {Marchegiani}},\ and\ \bibinfo
  {author} {\bibfnamefont {G.}~\bibnamefont {Germanese}},\ }\href@noop {}
  {\bibinfo {title} {Superconducting bipolar thermoelectric memory and method
  for writing a superconducting bipolar thermoelectric memory}},\ \bibinfo
  {howpublished} {Patent} (\bibinfo {year} {21/12/2021}),\ \bibinfo {note}
  {filing number: 102021000032042}\BibitemShut {NoStop}%
\bibitem [{\citenamefont {Germanese}\ \emph {et~al.}(2021)\citenamefont
  {Germanese}, \citenamefont {Paolucci}, \citenamefont {Marchegiani},
  \citenamefont {Braggio},\ and\ \citenamefont {Giazotto}}]{Germanese2021}%
  \BibitemOpen
  \bibfield  {author} {\bibinfo {author} {\bibfnamefont {G.}~\bibnamefont
  {Germanese}}, \bibinfo {author} {\bibfnamefont {F.}~\bibnamefont {Paolucci}},
  \bibinfo {author} {\bibfnamefont {G.}~\bibnamefont {Marchegiani}}, \bibinfo
  {author} {\bibfnamefont {A.}~\bibnamefont {Braggio}},\ and\ \bibinfo {author}
  {\bibfnamefont {F.}~\bibnamefont {Giazotto}},\ }\bibfield  {title} {\bibinfo
  {title} {Spontaneous symmetry breaking induced thermospin effect in
  superconducting tunnel junctions},\ }\href@noop {} {\bibfield  {journal}
  {\bibinfo  {journal} {Phys. Rev. B}\ }\textbf {\bibinfo {volume} {104}},\
  \bibinfo {pages} {184502} (\bibinfo {year} {2021})}\BibitemShut {NoStop}%
\bibitem [{\citenamefont {Guarcello}\ \emph {et~al.}(2019)\citenamefont
  {Guarcello}, \citenamefont {Braggio}, \citenamefont {Solinas},\ and\
  \citenamefont {Giazotto}}]{Guarcello1}%
  \BibitemOpen
  \bibfield  {author} {\bibinfo {author} {\bibfnamefont {C.}~\bibnamefont
  {Guarcello}}, \bibinfo {author} {\bibfnamefont {A.}~\bibnamefont {Braggio}},
  \bibinfo {author} {\bibfnamefont {P.}~\bibnamefont {Solinas}},\ and\ \bibinfo
  {author} {\bibfnamefont {F.}~\bibnamefont {Giazotto}},\ }\bibfield  {title}
  {\bibinfo {title} {Nonlinear critical-current thermal response of an
  asymmetric josephson tunnel junction},\ }\href@noop {} {\bibfield  {journal}
  {\bibinfo  {journal} {Phys. Rev. Applied}\ }\textbf {\bibinfo {volume}
  {11}},\ \bibinfo {pages} {024002} (\bibinfo {year} {2019})}\BibitemShut
  {NoStop}%
\bibitem [{\citenamefont {Guarcello}\ \emph {et~al.}(2022)\citenamefont
  {Guarcello}, \citenamefont {Citro}, \citenamefont {Giazotto},\ and\
  \citenamefont {Braggio}}]{Guarcello2}%
  \BibitemOpen
  \bibfield  {author} {\bibinfo {author} {\bibfnamefont {C.}~\bibnamefont
  {Guarcello}}, \bibinfo {author} {\bibfnamefont {R.}~\bibnamefont {Citro}},
  \bibinfo {author} {\bibfnamefont {F.}~\bibnamefont {Giazotto}},\ and\
  \bibinfo {author} {\bibfnamefont {A.}~\bibnamefont {Braggio}},\ }\bibfield
  {title} {\bibinfo {title} {Temperature-biased double-loop josephson flux
  transducer},\ }\href@noop {} {\bibfield  {journal} {\bibinfo  {journal}
  {Phys. Rev. Applied}\ }\textbf {\bibinfo {volume} {18}},\ \bibinfo {pages}
  {014037} (\bibinfo {year} {2022})}\BibitemShut {NoStop}%
\bibitem [{\citenamefont {Kittel}(1986)}]{KittelSS}%
  \BibitemOpen
  \bibfield  {author} {\bibinfo {author} {\bibfnamefont {C.}~\bibnamefont
  {Kittel}},\ }\href@noop {} {\emph {\bibinfo {title} {Introduction to Solid
  State Physics}}},\ \bibinfo {edition} {6th}\ ed.\ (\bibinfo  {publisher}
  {John Wiley \& sons, Inc. (New York)},\ \bibinfo {year} {1986})\BibitemShut
  {NoStop}%
\bibitem [{\citenamefont {Tinkham}(1996)}]{Tinkham}%
  \BibitemOpen
  \bibfield  {author} {\bibinfo {author} {\bibfnamefont {M.}~\bibnamefont
  {Tinkham}},\ }\href@noop {} {\emph {\bibinfo {title} {Introduction to
  superconductivity}}},\ \bibinfo {edition} {2nd}\ ed.\ (\bibinfo  {publisher}
  {McGraw-Hill},\ \bibinfo {year} {1996})\BibitemShut {NoStop}%
\bibitem [{\citenamefont {Bernazzani}(2021)}]{UniPiNonlinear}%
  \BibitemOpen
  \bibfield  {author} {\bibinfo {author} {\bibfnamefont {L.}~\bibnamefont
  {Bernazzani}},\ }\emph {\bibinfo {title} {Nonlinear Thermoelectricity in
  Nano-Devices}},\ \href@noop {} {Master's thesis},\ \bibinfo  {school}
  {University of Pisa} (\bibinfo {year} {2021})\BibitemShut {NoStop}%
\bibitem [{\citenamefont {Oostinga}\ \emph {et~al.}(2008)\citenamefont
  {Oostinga}, \citenamefont {Heersche}, \citenamefont {Liu}, \citenamefont
  {Morpurgo},\ and\ \citenamefont {Vandersypen}}]{NatureMaterials}%
  \BibitemOpen
  \bibfield  {author} {\bibinfo {author} {\bibfnamefont {J.~B.}\ \bibnamefont
  {Oostinga}}, \bibinfo {author} {\bibfnamefont {H.~B.}\ \bibnamefont
  {Heersche}}, \bibinfo {author} {\bibfnamefont {X.}~\bibnamefont {Liu}},
  \bibinfo {author} {\bibfnamefont {A.~F.}\ \bibnamefont {Morpurgo}},\ and\
  \bibinfo {author} {\bibfnamefont {L.~M.~K.}\ \bibnamefont {Vandersypen}},\
  }\bibfield  {title} {\bibinfo {title} {Gate-induced insulating state in
  bilayer graphene devices},\ }\href@noop {} {\bibfield  {journal} {\bibinfo
  {journal} {Nature Mater.}\ }\textbf {\bibinfo {volume} {7}},\ \bibinfo
  {pages} {151} (\bibinfo {year} {2008})}\BibitemShut {NoStop}%
\bibitem [{\citenamefont {Zhang}\ \emph {et~al.}(2009)\citenamefont {Zhang},
  \citenamefont {Tang}, \citenamefont {Girit}, \citenamefont {Hao},
  \citenamefont {Martin}, \citenamefont {Zettl}, \citenamefont {Crommie},
  \citenamefont {Shen},\ and\ \citenamefont {Wang}}]{NatureZhang}%
  \BibitemOpen
  \bibfield  {author} {\bibinfo {author} {\bibfnamefont {Y.}~\bibnamefont
  {Zhang}}, \bibinfo {author} {\bibfnamefont {T.}~\bibnamefont {Tang}},
  \bibinfo {author} {\bibfnamefont {C.}~\bibnamefont {Girit}}, \bibinfo
  {author} {\bibfnamefont {Z.}~\bibnamefont {Hao}}, \bibinfo {author}
  {\bibfnamefont {M.}~\bibnamefont {Martin}}, \bibinfo {author} {\bibfnamefont
  {A.}~\bibnamefont {Zettl}}, \bibinfo {author} {\bibfnamefont
  {M.}~\bibnamefont {Crommie}}, \bibinfo {author} {\bibfnamefont {Y.~R.}\
  \bibnamefont {Shen}},\ and\ \bibinfo {author} {\bibfnamefont
  {F.}~\bibnamefont {Wang}},\ }\bibfield  {title} {\bibinfo {title} {Direct
  observation of a widely tunable bandgap in bilayer graphene},\ }\href@noop {}
  {\bibfield  {journal} {\bibinfo  {journal} {Nature}\ }\textbf {\bibinfo
  {volume} {459}},\ \bibinfo {pages} {820} (\bibinfo {year}
  {2009})}\BibitemShut {NoStop}%
\bibitem [{\citenamefont {Castro}\ \emph {et~al.}(2010)\citenamefont {Castro},
  \citenamefont {Novoselov}, \citenamefont {Morozov}, \citenamefont {Peres},
  \citenamefont {dos Santos}, \citenamefont {Nilsson}, \citenamefont {Guinea},
  \citenamefont {Geim},\ and\ \citenamefont {Castro~Neto}}]{Castro1}%
  \BibitemOpen
  \bibfield  {author} {\bibinfo {author} {\bibfnamefont {E.~V.}\ \bibnamefont
  {Castro}}, \bibinfo {author} {\bibfnamefont {K.~S.}\ \bibnamefont
  {Novoselov}}, \bibinfo {author} {\bibfnamefont {S.~V.}\ \bibnamefont
  {Morozov}}, \bibinfo {author} {\bibfnamefont {N.~M.~R.}\ \bibnamefont
  {Peres}}, \bibinfo {author} {\bibfnamefont {J.~M. B.~L.}\ \bibnamefont {dos
  Santos}}, \bibinfo {author} {\bibfnamefont {J.}~\bibnamefont {Nilsson}},
  \bibinfo {author} {\bibfnamefont {F.}~\bibnamefont {Guinea}}, \bibinfo
  {author} {\bibfnamefont {A.~K.}\ \bibnamefont {Geim}},\ and\ \bibinfo
  {author} {\bibfnamefont {A.~H.}\ \bibnamefont {Castro~Neto}},\ }\bibfield
  {title} {\bibinfo {title} {Electronic properties of a biased graphene
  bilayer},\ }\href@noop {} {\bibfield  {journal} {\bibinfo  {journal} {J.
  Phys. Condens. Matter}\ }\textbf {\bibinfo {volume} {22}},\ \bibinfo {pages}
  {175503} (\bibinfo {year} {2010})}\BibitemShut {NoStop}%
\bibitem [{\citenamefont {McCann}\ and\ \citenamefont
  {Koshino}(2013)}]{McCann1}%
  \BibitemOpen
  \bibfield  {author} {\bibinfo {author} {\bibfnamefont {E.}~\bibnamefont
  {McCann}}\ and\ \bibinfo {author} {\bibfnamefont {M.}~\bibnamefont
  {Koshino}},\ }\bibfield  {title} {\bibinfo {title} {The electronic properties
  of bilayer graphene},\ }\href@noop {} {\bibfield  {journal} {\bibinfo
  {journal} {Rep. Prog. Phys.}\ }\textbf {\bibinfo {volume} {76}},\ \bibinfo
  {pages} {056503} (\bibinfo {year} {2013})}\BibitemShut {NoStop}%
\bibitem [{Not()}]{NoteNew}%
  \BibitemOpen
  \href@noop {} {}\bibinfo {note} {The lowest energy bands shown in the plot
  are obtained using Eq.~43 of Ref.~\cite{McCann1} and relating the potential
  $U$ to the bandgap 2$E_g$ with Eq.~\eqref{BLGGap}.}\BibitemShut {Stop}%
\bibitem [{\citenamefont {Balandin}(2011)}]{Baladin}%
  \BibitemOpen
  \bibfield  {author} {\bibinfo {author} {\bibfnamefont {A.}~\bibnamefont
  {Balandin}},\ }\bibfield  {title} {\bibinfo {title} {Thermal properties of
  graphene and nanostructured carbon materials},\ }\href@noop {} {\bibfield
  {journal} {\bibinfo  {journal} {Nature Mater.}\ }\textbf {\bibinfo {volume}
  {10}},\ \bibinfo {pages} {569} (\bibinfo {year} {2011})}\BibitemShut
  {NoStop}%
\bibitem [{\citenamefont {Dresselhaus}\ \emph {et~al.}(1999)\citenamefont
  {Dresselhaus}, \citenamefont {Dresselhaus}, \citenamefont {Sun},
  \citenamefont {Zhang}, \citenamefont {Cronin},\ and\ \citenamefont
  {Koga}}]{Dresselhaus1999}%
  \BibitemOpen
  \bibfield  {author} {\bibinfo {author} {\bibfnamefont {M.~S.}\ \bibnamefont
  {Dresselhaus}}, \bibinfo {author} {\bibfnamefont {G.}~\bibnamefont
  {Dresselhaus}}, \bibinfo {author} {\bibfnamefont {X.}~\bibnamefont {Sun}},
  \bibinfo {author} {\bibfnamefont {Z.}~\bibnamefont {Zhang}}, \bibinfo
  {author} {\bibfnamefont {S.~B.}\ \bibnamefont {Cronin}},\ and\ \bibinfo
  {author} {\bibfnamefont {T.}~\bibnamefont {Koga}},\ }\bibfield  {title}
  {\bibinfo {title} {Low-dimensional thermoelectric materials},\ }\href@noop {}
  {\bibfield  {journal} {\bibinfo  {journal} {Phys. Solid State}\ }\textbf
  {\bibinfo {volume} {41}},\ \bibinfo {pages} {679} (\bibinfo {year}
  {1999})}\BibitemShut {NoStop}%
\bibitem [{\citenamefont {Novoselov}\ \emph {et~al.}(2005)\citenamefont
  {Novoselov}, \citenamefont {Geim}, \citenamefont {Morozov}, \citenamefont
  {Jiang}, \citenamefont {Katsnelson}, \citenamefont {Grigorieva},
  \citenamefont {Dubonos},\ and\ \citenamefont {Firsov}}]{Novoselov2005}%
  \BibitemOpen
  \bibfield  {author} {\bibinfo {author} {\bibfnamefont {K.}~\bibnamefont
  {Novoselov}}, \bibinfo {author} {\bibfnamefont {A.}~\bibnamefont {Geim}},
  \bibinfo {author} {\bibfnamefont {S.}~\bibnamefont {Morozov}}, \bibinfo
  {author} {\bibfnamefont {D.}~\bibnamefont {Jiang}}, \bibinfo {author}
  {\bibfnamefont {M.}~\bibnamefont {Katsnelson}}, \bibinfo {author}
  {\bibfnamefont {I.~V.}\ \bibnamefont {Grigorieva}}, \bibinfo {author}
  {\bibfnamefont {S.~V.}\ \bibnamefont {Dubonos}},\ and\ \bibinfo {author}
  {\bibfnamefont {A.~A.}\ \bibnamefont {Firsov}},\ }\bibfield  {title}
  {\bibinfo {title} {Two-dimensional gas of massless dirac fermions in
  graphene},\ }\href@noop {} {\bibfield  {journal} {\bibinfo  {journal}
  {Nature}\ }\textbf {\bibinfo {volume} {438}},\ \bibinfo {pages} {197}
  (\bibinfo {year} {2005})}\BibitemShut {NoStop}%
\bibitem [{\citenamefont {Sprinkle}\ \emph {et~al.}(2009)\citenamefont
  {Sprinkle}, \citenamefont {Siegel}, \citenamefont {Hu}, \citenamefont
  {Hicks}, \citenamefont {Tejeda}, \citenamefont {Taleb-Ibrahimi},
  \citenamefont {Le~F\`evre}, \citenamefont {Bertran}, \citenamefont {Vizzini},
  \citenamefont {Enriquez}, \citenamefont {Chiang}, \citenamefont
  {Soukiassian}, \citenamefont {Berger}, \citenamefont {de~Heer}, \citenamefont
  {Lanzara},\ and\ \citenamefont {Conrad}}]{Sprinkle2009}%
  \BibitemOpen
  \bibfield  {author} {\bibinfo {author} {\bibfnamefont {M.}~\bibnamefont
  {Sprinkle}}, \bibinfo {author} {\bibfnamefont {D.}~\bibnamefont {Siegel}},
  \bibinfo {author} {\bibfnamefont {Y.}~\bibnamefont {Hu}}, \bibinfo {author}
  {\bibfnamefont {J.}~\bibnamefont {Hicks}}, \bibinfo {author} {\bibfnamefont
  {A.}~\bibnamefont {Tejeda}}, \bibinfo {author} {\bibfnamefont
  {A.}~\bibnamefont {Taleb-Ibrahimi}}, \bibinfo {author} {\bibfnamefont
  {P.}~\bibnamefont {Le~F\`evre}}, \bibinfo {author} {\bibfnamefont
  {F.}~\bibnamefont {Bertran}}, \bibinfo {author} {\bibfnamefont
  {S.}~\bibnamefont {Vizzini}}, \bibinfo {author} {\bibfnamefont
  {H.}~\bibnamefont {Enriquez}}, \bibinfo {author} {\bibfnamefont
  {S.}~\bibnamefont {Chiang}}, \bibinfo {author} {\bibfnamefont
  {P.}~\bibnamefont {Soukiassian}}, \bibinfo {author} {\bibfnamefont
  {C.}~\bibnamefont {Berger}}, \bibinfo {author} {\bibfnamefont {W.~A.}\
  \bibnamefont {de~Heer}}, \bibinfo {author} {\bibfnamefont {A.}~\bibnamefont
  {Lanzara}},\ and\ \bibinfo {author} {\bibfnamefont {E.~H.}\ \bibnamefont
  {Conrad}},\ }\bibfield  {title} {\bibinfo {title} {First direct observation
  of a nearly ideal graphene band structure},\ }\href@noop {} {\bibfield
  {journal} {\bibinfo  {journal} {Phys. Rev. Lett.}\ }\textbf {\bibinfo
  {volume} {103}},\ \bibinfo {pages} {226803} (\bibinfo {year}
  {2009})}\BibitemShut {NoStop}%
\bibitem [{\citenamefont {McCann}(2006)}]{McCann2}%
  \BibitemOpen
  \bibfield  {author} {\bibinfo {author} {\bibfnamefont {E.}~\bibnamefont
  {McCann}},\ }\bibfield  {title} {\bibinfo {title} {Asymmetry gap in the
  electronic band structure of bilayer graphene},\ }\href@noop {} {\bibfield
  {journal} {\bibinfo  {journal} {Phys. Rev. B}\ }\textbf {\bibinfo {volume}
  {74}},\ \bibinfo {pages} {161403} (\bibinfo {year} {2006})}\BibitemShut
  {NoStop}%
\bibitem [{\citenamefont {S\l{}awi\ifmmode~\acute{n}\else \'{n}\fi{}ska}\ \emph
  {et~al.}(2010)\citenamefont {S\l{}awi\ifmmode~\acute{n}\else \'{n}\fi{}ska},
  \citenamefont {Zasada},\ and\ \citenamefont {Klusek}}]{Slavinska}%
  \BibitemOpen
  \bibfield  {author} {\bibinfo {author} {\bibfnamefont {J.}~\bibnamefont
  {S\l{}awi\ifmmode~\acute{n}\else \'{n}\fi{}ska}}, \bibinfo {author}
  {\bibfnamefont {I.}~\bibnamefont {Zasada}},\ and\ \bibinfo {author}
  {\bibfnamefont {Z.}~\bibnamefont {Klusek}},\ }\bibfield  {title} {\bibinfo
  {title} {Energy gap tuning in graphene on hexagonal boron nitride bilayer
  system},\ }\href@noop {} {\bibfield  {journal} {\bibinfo  {journal} {Phys.
  Rev. B}\ }\textbf {\bibinfo {volume} {81}},\ \bibinfo {pages} {155433}
  (\bibinfo {year} {2010})}\BibitemShut {NoStop}%
\bibitem [{\citenamefont {Mak}\ \emph {et~al.}(2009)\citenamefont {Mak},
  \citenamefont {Lui}, \citenamefont {Shan},\ and\ \citenamefont
  {Heinz}}]{Mak}%
  \BibitemOpen
  \bibfield  {author} {\bibinfo {author} {\bibfnamefont {K.~F.}\ \bibnamefont
  {Mak}}, \bibinfo {author} {\bibfnamefont {C.~H.}\ \bibnamefont {Lui}},
  \bibinfo {author} {\bibfnamefont {J.}~\bibnamefont {Shan}},\ and\ \bibinfo
  {author} {\bibfnamefont {T.~F.}\ \bibnamefont {Heinz}},\ }\bibfield  {title}
  {\bibinfo {title} {Observation of an electric-field-induced band gap in
  bilayer graphene by infrared spectroscopy},\ }\href@noop {} {\bibfield
  {journal} {\bibinfo  {journal} {Phys. Rev. Lett.}\ }\textbf {\bibinfo
  {volume} {102}},\ \bibinfo {pages} {256405} (\bibinfo {year}
  {2009})}\BibitemShut {NoStop}%
\bibitem [{\citenamefont {Vischi}\ \emph {et~al.}(2020)\citenamefont {Vischi},
  \citenamefont {Carrega}, \citenamefont {Braggio}, \citenamefont {Paolucci},
  \citenamefont {Bianco}, \citenamefont {Roddaro},\ and\ \citenamefont
  {Giazotto}}]{Vischi}%
  \BibitemOpen
  \bibfield  {author} {\bibinfo {author} {\bibfnamefont {F.}~\bibnamefont
  {Vischi}}, \bibinfo {author} {\bibfnamefont {M.}~\bibnamefont {Carrega}},
  \bibinfo {author} {\bibfnamefont {A.}~\bibnamefont {Braggio}}, \bibinfo
  {author} {\bibfnamefont {F.}~\bibnamefont {Paolucci}}, \bibinfo {author}
  {\bibfnamefont {F.}~\bibnamefont {Bianco}}, \bibinfo {author} {\bibfnamefont
  {S.}~\bibnamefont {Roddaro}},\ and\ \bibinfo {author} {\bibfnamefont
  {F.}~\bibnamefont {Giazotto}},\ }\bibfield  {title} {\bibinfo {title}
  {Electron cooling with graphene-insulator-superconductor tunnel junctions for
  applications in fast bolometry},\ }\href@noop {} {\bibfield  {journal}
  {\bibinfo  {journal} {Phys. Rev. Applied}\ }\textbf {\bibinfo {volume}
  {13}},\ \bibinfo {pages} {054006} (\bibinfo {year} {2020})}\BibitemShut
  {NoStop}%
\bibitem [{\citenamefont {Viljas}\ and\ \citenamefont
  {Heikkil\"a}(2010)}]{Viljas}%
  \BibitemOpen
  \bibfield  {author} {\bibinfo {author} {\bibfnamefont {J.~K.}\ \bibnamefont
  {Viljas}}\ and\ \bibinfo {author} {\bibfnamefont {T.~T.}\ \bibnamefont
  {Heikkil\"a}},\ }\bibfield  {title} {\bibinfo {title} {Electron-phonon heat
  transfer in monolayer and bilayer graphene},\ }\href@noop {} {\bibfield
  {journal} {\bibinfo  {journal} {Phys. Rev. B}\ }\textbf {\bibinfo {volume}
  {81}},\ \bibinfo {pages} {245404} (\bibinfo {year} {2010})}\BibitemShut
  {NoStop}%
\bibitem [{Note1()}]{Note1}%
  \BibitemOpen
  \bibinfo {note} {The top/bottom gates are referred to the BLG Ohmic contacts
  in order to induce a proper gating on the BLG which is not affected by the
  bias applied to the junction.}\BibitemShut {Stop}%
\bibitem [{\citenamefont {Maestre}\ \emph {et~al.}(2021)\citenamefont
  {Maestre}, \citenamefont {Toury}, \citenamefont {Steyer}, \citenamefont
  {Garnier},\ and\ \citenamefont {Journet}}]{Maestre2021}%
  \BibitemOpen
  \bibfield  {author} {\bibinfo {author} {\bibfnamefont {C.}~\bibnamefont
  {Maestre}}, \bibinfo {author} {\bibfnamefont {B.}~\bibnamefont {Toury}},
  \bibinfo {author} {\bibfnamefont {P.}~\bibnamefont {Steyer}}, \bibinfo
  {author} {\bibfnamefont {V.}~\bibnamefont {Garnier}},\ and\ \bibinfo {author}
  {\bibfnamefont {C.}~\bibnamefont {Journet}},\ }\bibfield  {title} {\bibinfo
  {title} {Hexagonal boron nitride: a review on selfstanding crystals synthesis
  towards 2d nanosheets},\ }\href@noop {} {\bibfield  {journal} {\bibinfo
  {journal} {J. Phys. Mater.}\ }\textbf {\bibinfo {volume} {4}},\ \bibinfo
  {pages} {044018} (\bibinfo {year} {2021})}\BibitemShut {NoStop}%
\bibitem [{\citenamefont {Tinkham}(1972)}]{Tinkham1972}%
  \BibitemOpen
  \bibfield  {author} {\bibinfo {author} {\bibfnamefont {M.}~\bibnamefont
  {Tinkham}},\ }\bibfield  {title} {\bibinfo {title} {Tunneling generation,
  relaxation, and tunneling detection of hole-electron imbalance in
  superconductors},\ }\href@noop {} {\bibfield  {journal} {\bibinfo  {journal}
  {Phys. Rev. B}\ }\textbf {\bibinfo {volume} {6}},\ \bibinfo {pages} {1747}
  (\bibinfo {year} {1972})}\BibitemShut {NoStop}%
\bibitem [{\citenamefont {Ketterson}\ and\ \citenamefont
  {Song}(1999)}]{Ketterson}%
  \BibitemOpen
  \bibfield  {author} {\bibinfo {author} {\bibfnamefont {J.~B.}\ \bibnamefont
  {Ketterson}}\ and\ \bibinfo {author} {\bibfnamefont {S.~N.}\ \bibnamefont
  {Song}},\ }\href@noop {} {\emph {\bibinfo {title} {Superconductivity}}}\
  (\bibinfo  {publisher} {Cambridge University Press},\ \bibinfo {year}
  {1999})\BibitemShut {NoStop}%
\bibitem [{\citenamefont {Giazotto}\ \emph {et~al.}(2006)\citenamefont
  {Giazotto}, \citenamefont {Heikkil\"a}, \citenamefont {Luukanen},
  \citenamefont {Savin},\ and\ \citenamefont {Pekola}}]{RMPGiazotto}%
  \BibitemOpen
  \bibfield  {author} {\bibinfo {author} {\bibfnamefont {F.}~\bibnamefont
  {Giazotto}}, \bibinfo {author} {\bibfnamefont {T.~T.}\ \bibnamefont
  {Heikkil\"a}}, \bibinfo {author} {\bibfnamefont {A.}~\bibnamefont
  {Luukanen}}, \bibinfo {author} {\bibfnamefont {A.}~\bibnamefont {Savin}},\
  and\ \bibinfo {author} {\bibfnamefont {J.~P.}\ \bibnamefont {Pekola}},\
  }\bibfield  {title} {\bibinfo {title} {Opportunities for mesoscopics in
  thermometry and refrigeration: Physics and applications},\ }\href@noop {}
  {\bibfield  {journal} {\bibinfo  {journal} {Rev. Mod. Phys.}\ }\textbf
  {\bibinfo {volume} {78}},\ \bibinfo {pages} {217} (\bibinfo {year}
  {2006})}\BibitemShut {NoStop}%
\bibitem [{\citenamefont {Dynes}\ \emph {et~al.}(1984)\citenamefont {Dynes},
  \citenamefont {Garno}, \citenamefont {Hertel},\ and\ \citenamefont
  {Orlando}}]{Dynes}%
  \BibitemOpen
  \bibfield  {author} {\bibinfo {author} {\bibfnamefont {R.~C.}\ \bibnamefont
  {Dynes}}, \bibinfo {author} {\bibfnamefont {J.~P.}\ \bibnamefont {Garno}},
  \bibinfo {author} {\bibfnamefont {G.~B.}\ \bibnamefont {Hertel}},\ and\
  \bibinfo {author} {\bibfnamefont {T.~P.}\ \bibnamefont {Orlando}},\
  }\bibfield  {title} {\bibinfo {title} {Tunneling study of superconductivity
  near the metal-insulator transition},\ }\href@noop {} {\bibfield  {journal}
  {\bibinfo  {journal} {Phys. Rev. Lett.}\ }\textbf {\bibinfo {volume} {53}},\
  \bibinfo {pages} {2437} (\bibinfo {year} {1984})}\BibitemShut {NoStop}%
\bibitem [{\citenamefont {Su\'arez~Morell}\ and\ \citenamefont
  {Foa~Torres}(2012)}]{Suarez}%
  \BibitemOpen
  \bibfield  {author} {\bibinfo {author} {\bibfnamefont {E.}~\bibnamefont
  {Su\'arez~Morell}}\ and\ \bibinfo {author} {\bibfnamefont {L.~E.~F.}\
  \bibnamefont {Foa~Torres}},\ }\bibfield  {title} {\bibinfo {title} {Radiation
  effects on the electronic properties of bilayer graphene},\ }\href@noop {}
  {\bibfield  {journal} {\bibinfo  {journal} {Phys. Rev. B}\ }\textbf {\bibinfo
  {volume} {86}},\ \bibinfo {pages} {125449} (\bibinfo {year}
  {2012})}\BibitemShut {NoStop}%
\bibitem [{Note2()}]{Note2}%
  \BibitemOpen
  \bibinfo {note} {The hBN is commonly used to encapsulate graphene since its
  crystalline configuration matches well with that of graphene diminishing the
  surface disorder at the interfaces.}\BibitemShut {Stop}%
\bibitem [{Note3()}]{Note3}%
  \BibitemOpen
  \bibinfo {note} {The conductance $G_T=4 N_Fe^2|\protect \mathcal
  {T}|^2t_\perp /(\hbar \protect \sqrt {3}t^2)$ would represent the
  differential conductance of the tunnel junction when the superconductor is in
  the normal state and the BLG is gapless.}\BibitemShut {Stop}%
\bibitem [{\citenamefont {Britnell}\ \emph {et~al.}(2012)\citenamefont
  {Britnell}, \citenamefont {Gorbachev}, \citenamefont {Jalil}, \citenamefont
  {Belle}, \citenamefont {Schedin}, \citenamefont {Katsnelson}, \citenamefont
  {Eaves}, \citenamefont {Morozov}, \citenamefont {Mayorov}, \citenamefont
  {Peres}, \citenamefont {Castro~Neto}, \citenamefont {Leist}, \citenamefont
  {Geim}, \citenamefont {Ponomarenko},\ and\ \citenamefont
  {Novoselov}}]{NanoLetters}%
  \BibitemOpen
  \bibfield  {author} {\bibinfo {author} {\bibfnamefont {L.}~\bibnamefont
  {Britnell}}, \bibinfo {author} {\bibfnamefont {R.~V.}\ \bibnamefont
  {Gorbachev}}, \bibinfo {author} {\bibfnamefont {R.}~\bibnamefont {Jalil}},
  \bibinfo {author} {\bibfnamefont {B.~D.}\ \bibnamefont {Belle}}, \bibinfo
  {author} {\bibfnamefont {F.}~\bibnamefont {Schedin}}, \bibinfo {author}
  {\bibfnamefont {M.~I.}\ \bibnamefont {Katsnelson}}, \bibinfo {author}
  {\bibfnamefont {L.}~\bibnamefont {Eaves}}, \bibinfo {author} {\bibfnamefont
  {S.~V.}\ \bibnamefont {Morozov}}, \bibinfo {author} {\bibfnamefont {A.~S.}\
  \bibnamefont {Mayorov}}, \bibinfo {author} {\bibfnamefont {N.~M.~R.}\
  \bibnamefont {Peres}}, \bibinfo {author} {\bibfnamefont {A.~H.}\ \bibnamefont
  {Castro~Neto}}, \bibinfo {author} {\bibfnamefont {J.}~\bibnamefont {Leist}},
  \bibinfo {author} {\bibfnamefont {A.~K.}\ \bibnamefont {Geim}}, \bibinfo
  {author} {\bibfnamefont {L.~A.}\ \bibnamefont {Ponomarenko}},\ and\ \bibinfo
  {author} {\bibfnamefont {K.~S.}\ \bibnamefont {Novoselov}},\ }\bibfield
  {title} {\bibinfo {title} {Electron tunneling through ultrathin boron nitride
  crystalline barriers},\ }\href@noop {} {\bibfield  {journal} {\bibinfo
  {journal} {Nano Lett.}\ }\textbf {\bibinfo {volume} {12}},\ \bibinfo {pages}
  {1707} (\bibinfo {year} {2012})}\BibitemShut {NoStop}%
\bibitem [{\citenamefont {Ramezani}\ \emph {et~al.}(2021)\citenamefont
  {Ramezani}, \citenamefont {Sampaio}, \citenamefont {Watanabe}, \citenamefont
  {Taniguchi}, \citenamefont {Schönenberger},\ and\ \citenamefont
  {Baumgartner}}]{PinHole}%
  \BibitemOpen
  \bibfield  {author} {\bibinfo {author} {\bibfnamefont {M.}~\bibnamefont
  {Ramezani}}, \bibinfo {author} {\bibfnamefont {I.}~\bibnamefont {Sampaio}},
  \bibinfo {author} {\bibfnamefont {K.}~\bibnamefont {Watanabe}}, \bibinfo
  {author} {\bibfnamefont {T.}~\bibnamefont {Taniguchi}}, \bibinfo {author}
  {\bibfnamefont {C.}~\bibnamefont {Schönenberger}},\ and\ \bibinfo {author}
  {\bibfnamefont {A.}~\bibnamefont {Baumgartner}},\ }\bibfield  {title}
  {\bibinfo {title} {Superconducting contacts to a monolayer semiconductor},\
  }\href@noop {} {\bibfield  {journal} {\bibinfo  {journal} {Nano Lett.}\
  }\textbf {\bibinfo {volume} {21}},\ \bibinfo {pages} {5614} (\bibinfo {year}
  {2021})}\BibitemShut {NoStop}%
\bibitem [{\citenamefont {Padilha}\ \emph {et~al.}(2012)\citenamefont
  {Padilha}, \citenamefont {Pontes},\ and\ \citenamefont {Fazzio}}]{Padilha}%
  \BibitemOpen
  \bibfield  {author} {\bibinfo {author} {\bibfnamefont {J.~E.}\ \bibnamefont
  {Padilha}}, \bibinfo {author} {\bibfnamefont {R.~B.}\ \bibnamefont
  {Pontes}},\ and\ \bibinfo {author} {\bibfnamefont {A.}~\bibnamefont
  {Fazzio}},\ }\bibfield  {title} {\bibinfo {title} {Bilayer graphene on h-bn
  substrate: investigating the breakdown voltage and tuning the bandgap by
  electric field},\ }\href@noop {} {\bibfield  {journal} {\bibinfo  {journal}
  {J. Phys. Condens. Matter}\ }\textbf {\bibinfo {volume} {24}},\ \bibinfo
  {pages} {075301} (\bibinfo {year} {2012})}\BibitemShut {NoStop}%
\bibitem [{\citenamefont {Castro~Neto}\ \emph {et~al.}(2009)\citenamefont
  {Castro~Neto}, \citenamefont {Guinea}, \citenamefont {Peres}, \citenamefont
  {Novoselov},\ and\ \citenamefont {Geim}}]{Castro2}%
  \BibitemOpen
  \bibfield  {author} {\bibinfo {author} {\bibfnamefont {A.~H.}\ \bibnamefont
  {Castro~Neto}}, \bibinfo {author} {\bibfnamefont {F.}~\bibnamefont {Guinea}},
  \bibinfo {author} {\bibfnamefont {N.~M.~R.}\ \bibnamefont {Peres}}, \bibinfo
  {author} {\bibfnamefont {K.~S.}\ \bibnamefont {Novoselov}},\ and\ \bibinfo
  {author} {\bibfnamefont {A.~K.}\ \bibnamefont {Geim}},\ }\bibfield  {title}
  {\bibinfo {title} {The electronic properties of graphene},\ }\href@noop {}
  {\bibfield  {journal} {\bibinfo  {journal} {Rev. Mod. Phys.}\ }\textbf
  {\bibinfo {volume} {81}},\ \bibinfo {pages} {109} (\bibinfo {year}
  {2009})}\BibitemShut {NoStop}%
\bibitem [{\citenamefont {Dubi}\ and\ \citenamefont
  {Di~Ventra}(2011)}]{DiVentra}%
  \BibitemOpen
  \bibfield  {author} {\bibinfo {author} {\bibfnamefont {Y.}~\bibnamefont
  {Dubi}}\ and\ \bibinfo {author} {\bibfnamefont {M.}~\bibnamefont
  {Di~Ventra}},\ }\bibfield  {title} {\bibinfo {title} {Colloquium: Heat flow
  and thermoelectricity in atomic and molecular junctions},\ }\href@noop {}
  {\bibfield  {journal} {\bibinfo  {journal} {Rev. Mod. Phys.}\ }\textbf
  {\bibinfo {volume} {83}},\ \bibinfo {pages} {131} (\bibinfo {year}
  {2011})}\BibitemShut {NoStop}%
\bibitem [{Note4()}]{Note4}%
  \BibitemOpen
  \bibinfo {note} {In principle inverting the ratio between $E_g/\Delta _{0,R}$
  also the opposite temperature gradient may be assumed. But this is less
  practical since the temperature of the superconducting lead should not exceed
  its critical temperature in any case.}\BibitemShut {Stop}%
\bibitem [{\citenamefont {Kheradsoud}\ \emph {et~al.}(2019)\citenamefont
  {Kheradsoud}, \citenamefont {Dashti}, \citenamefont {Misiorny}, \citenamefont
  {Potts}, \citenamefont {Splettstoesser},\ and\ \citenamefont
  {Samuelsson}}]{Splett}%
  \BibitemOpen
  \bibfield  {author} {\bibinfo {author} {\bibfnamefont {S.}~\bibnamefont
  {Kheradsoud}}, \bibinfo {author} {\bibfnamefont {N.}~\bibnamefont {Dashti}},
  \bibinfo {author} {\bibfnamefont {M.}~\bibnamefont {Misiorny}}, \bibinfo
  {author} {\bibfnamefont {P.~P.}\ \bibnamefont {Potts}}, \bibinfo {author}
  {\bibfnamefont {J.}~\bibnamefont {Splettstoesser}},\ and\ \bibinfo {author}
  {\bibfnamefont {P.}~\bibnamefont {Samuelsson}},\ }\bibfield  {title}
  {\bibinfo {title} {Power, efficiency and fluctuations in a quantum point
  contact as steady-state thermoelectric heat engine},\ }\href@noop {}
  {\bibfield  {journal} {\bibinfo  {journal} {Entropy}\ }\textbf {\bibinfo
  {volume} {21}},\ \bibinfo {pages} {777} (\bibinfo {year} {2019})}\BibitemShut
  {NoStop}%
\bibitem [{Note5()}]{Note5}%
  \BibitemOpen
  \bibinfo {note} {Even if the absolute value of those quantity depends also on
  the Dynes' parameter $\Gamma _D$ the general behaviour would be quite
  independent on it.}\BibitemShut {Stop}%
\bibitem [{Note6()}]{Note6}%
  \BibitemOpen
  \bibinfo {note} {Obviously the heat may potentially be transferred also by
  other contributions such as phonon or even radiative contributions that are
  more difficult to be estimated. Our computation of efficiency represents a
  sort of optimal electronic efficiency of the device.}\BibitemShut {Stop}%
\bibitem [{\citenamefont {Alymov}\ \emph {et~al.}(2016)\citenamefont {Alymov},
  \citenamefont {Vyurkov}, \citenamefont {Ryzhii},\ and\ \citenamefont
  {Svintsov}}]{SciRep}%
  \BibitemOpen
  \bibfield  {author} {\bibinfo {author} {\bibfnamefont {G.}~\bibnamefont
  {Alymov}}, \bibinfo {author} {\bibfnamefont {V.}~\bibnamefont {Vyurkov}},
  \bibinfo {author} {\bibfnamefont {V.}~\bibnamefont {Ryzhii}},\ and\ \bibinfo
  {author} {\bibfnamefont {D.}~\bibnamefont {Svintsov}},\ }\bibfield  {title}
  {\bibinfo {title} {Abrupt current switching in graphene bilayer tunnel
  transistors enabled by van hove singularities},\ }\href@noop {} {\bibfield
  {journal} {\bibinfo  {journal} {Sci. Rep.}\ }\textbf {\bibinfo {volume}
  {6}},\ \bibinfo {pages} {24654} (\bibinfo {year} {2016})}\BibitemShut
  {NoStop}%
\bibitem [{\citenamefont {Icking}\ \emph {et~al.}(2022)\citenamefont {Icking},
  \citenamefont {Banszerus}, \citenamefont {Wörtche}, \citenamefont {Volmer},
  \citenamefont {Schmidt}, \citenamefont {Steiner}, \citenamefont {Engels},
  \citenamefont {Hesselmann}, \citenamefont {Goldsche}, \citenamefont
  {Watanabe}, \citenamefont {Taniguchi}, \citenamefont {Volk}, \citenamefont
  {Beschoten},\ and\ \citenamefont {Stampfer}}]{Icking2022}%
  \BibitemOpen
  \bibfield  {author} {\bibinfo {author} {\bibfnamefont {E.}~\bibnamefont
  {Icking}}, \bibinfo {author} {\bibfnamefont {L.}~\bibnamefont {Banszerus}},
  \bibinfo {author} {\bibfnamefont {F.}~\bibnamefont {Wörtche}}, \bibinfo
  {author} {\bibfnamefont {F.}~\bibnamefont {Volmer}}, \bibinfo {author}
  {\bibfnamefont {P.}~\bibnamefont {Schmidt}}, \bibinfo {author} {\bibfnamefont
  {C.}~\bibnamefont {Steiner}}, \bibinfo {author} {\bibfnamefont
  {S.}~\bibnamefont {Engels}}, \bibinfo {author} {\bibfnamefont
  {J.}~\bibnamefont {Hesselmann}}, \bibinfo {author} {\bibfnamefont
  {M.}~\bibnamefont {Goldsche}}, \bibinfo {author} {\bibfnamefont
  {K.}~\bibnamefont {Watanabe}}, \bibinfo {author} {\bibfnamefont
  {T.}~\bibnamefont {Taniguchi}}, \bibinfo {author} {\bibfnamefont
  {C.}~\bibnamefont {Volk}}, \bibinfo {author} {\bibfnamefont {B.}~\bibnamefont
  {Beschoten}},\ and\ \bibinfo {author} {\bibfnamefont {C.}~\bibnamefont
  {Stampfer}},\ }\bibfield  {title} {\bibinfo {title} {Transport spectroscopy
  of ultraclean tunable band gaps in bilayer graphene},\ }\href
  {https://onlinelibrary.wiley.com/doi/abs/10.1002/aelm.202200510} {\bibfield
  {journal} {\bibinfo  {journal} {Adv. Electron. Mater.}\ }\textbf {\bibinfo
  {volume} {8}},\ \bibinfo {pages} {2200510} (\bibinfo {year}
  {2022})}\BibitemShut {NoStop}%
\bibitem [{Note7()}]{Note7}%
  \BibitemOpen
  \bibinfo {note} {The presence of localized charge states induced by
  impurities also determines a modification of the gating conditions anyway
  these contributions can be simply accounted by including appropriate offsets
  to the gate voltages, so we neglected them~\cite {Icking2022}.}\BibitemShut
  {Stop}%
\bibitem [{Note8()}]{Note8}%
  \BibitemOpen
  \bibinfo {note} {The bandgap self-consistent expression of Eq.~\protect
  \eqref {eq:Uself} is derived using Eqs.~(65) and (74) of Ref.~\cite {McCann1}
  in the limit with $\varepsilon _r\sim 1$, $\Lambda \sim 1$ and assuming no
  mobile charge density when the Fermi energy is in the gap. Correspondingly
  the Fermi-energy expression Eq.~\protect \eqref {eq:FermiEnergy} coincides
  with the chemical potential of Eq.~(S13) of the supplementary information of
  Ref.~\cite {Icking2022} where have been neglected terms involving the inverse
  of BLG capacitance $C_{BLG}^{-1}$ being $C_{BLG}\gg C_b, C_t$.}\BibitemShut
  {Stop}%
\bibitem [{Note9()}]{Note9}%
  \BibitemOpen
  \bibinfo {note} {This simplified model may not be fully self-consistent when
  the system is gated well outside the gap, but corrections are expected to
  slightly renormalize the gap value not affecting physics
  discussed.}\BibitemShut {Stop}%
\bibitem [{\citenamefont {Paolucci}\ \emph {et~al.}(2023)\citenamefont
  {Paolucci}, \citenamefont {Germanese}, \citenamefont {Braggio},\ and\
  \citenamefont {Giazotto}}]{paolucci2023}%
  \BibitemOpen
  \bibfield  {author} {\bibinfo {author} {\bibfnamefont {F.}~\bibnamefont
  {Paolucci}}, \bibinfo {author} {\bibfnamefont {G.}~\bibnamefont {Germanese}},
  \bibinfo {author} {\bibfnamefont {A.}~\bibnamefont {Braggio}},\ and\ \bibinfo
  {author} {\bibfnamefont {F.}~\bibnamefont {Giazotto}},\ }\bibfield  {title}
  {\bibinfo {title} {A highly-sensitive broadband superconducting
  thermoelectric single-photon detector},\ }\href@noop {} {\bibfield  {journal}
  {\bibinfo  {journal} {arXiv preprint arXiv:2302.02933}\ } (\bibinfo {year}
  {2023})}\BibitemShut {NoStop}%
\bibitem [{\citenamefont {Xia}\ \emph {et~al.}(2009)\citenamefont {Xia},
  \citenamefont {Mueller}, \citenamefont {Lin}, \citenamefont {Valdes-Garcia},\
  and\ \citenamefont {Avouris}}]{Xia2009}%
  \BibitemOpen
  \bibfield  {author} {\bibinfo {author} {\bibfnamefont {F.}~\bibnamefont
  {Xia}}, \bibinfo {author} {\bibfnamefont {T.}~\bibnamefont {Mueller}},
  \bibinfo {author} {\bibfnamefont {Y.}~\bibnamefont {Lin}}, \bibinfo {author}
  {\bibfnamefont {A.}~\bibnamefont {Valdes-Garcia}},\ and\ \bibinfo {author}
  {\bibfnamefont {P.}~\bibnamefont {Avouris}},\ }\bibfield  {title} {\bibinfo
  {title} {Ultrafast graphene photodetector},\ }\href@noop {} {\bibfield
  {journal} {\bibinfo  {journal} {Nature Nanotech.}\ }\textbf {\bibinfo
  {volume} {4}},\ \bibinfo {pages} {839} (\bibinfo {year} {2009})}\BibitemShut
  {NoStop}%
\bibitem [{\citenamefont {Mueller}\ \emph {et~al.}(2010)\citenamefont
  {Mueller}, \citenamefont {Xia},\ and\ \citenamefont {Avouris}}]{Mueller2010}%
  \BibitemOpen
  \bibfield  {author} {\bibinfo {author} {\bibfnamefont {T.}~\bibnamefont
  {Mueller}}, \bibinfo {author} {\bibfnamefont {F.}~\bibnamefont {Xia}},\ and\
  \bibinfo {author} {\bibfnamefont {P.}~\bibnamefont {Avouris}},\ }\bibfield
  {title} {\bibinfo {title} {Graphene photodetectors for high-speed optical
  communications},\ }\href@noop {} {\bibfield  {journal} {\bibinfo  {journal}
  {Nature Photon.}\ }\textbf {\bibinfo {volume} {4}},\ \bibinfo {pages} {297}
  (\bibinfo {year} {2010})}\BibitemShut {NoStop}%
\bibitem [{\citenamefont {Lee}\ \emph {et~al.}(2020)\citenamefont {Lee},
  \citenamefont {Efetov}, \citenamefont {Jung}, \citenamefont {Ranzani},
  \citenamefont {Walsh}, \citenamefont {Okhi}, \citenamefont {Taniguchi},
  \citenamefont {Watanabe}, \citenamefont {Englund},\ and\ \citenamefont
  {Fong}}]{Lee2020}%
  \BibitemOpen
  \bibfield  {author} {\bibinfo {author} {\bibfnamefont {G.~H.}\ \bibnamefont
  {Lee}}, \bibinfo {author} {\bibfnamefont {D.~K.}\ \bibnamefont {Efetov}},
  \bibinfo {author} {\bibfnamefont {W.}~\bibnamefont {Jung}}, \bibinfo {author}
  {\bibfnamefont {L.}~\bibnamefont {Ranzani}}, \bibinfo {author} {\bibfnamefont
  {E.~D.}\ \bibnamefont {Walsh}}, \bibinfo {author} {\bibfnamefont {T.~A.}\
  \bibnamefont {Okhi}}, \bibinfo {author} {\bibfnamefont {T.}~\bibnamefont
  {Taniguchi}}, \bibinfo {author} {\bibfnamefont {K.}~\bibnamefont {Watanabe}},
  \bibinfo {author} {\bibfnamefont {P.~K.~D.}\ \bibnamefont {Englund}},\ and\
  \bibinfo {author} {\bibfnamefont {K.~C.}\ \bibnamefont {Fong}},\ }\bibfield
  {title} {\bibinfo {title} {Graphene-based josephson junction microwave
  bolometer},\ }\href@noop {} {\bibfield  {journal} {\bibinfo  {journal}
  {Nature}\ }\textbf {\bibinfo {volume} {586}},\ \bibinfo {pages} {42}
  (\bibinfo {year} {2020})}\BibitemShut {NoStop}%
\bibitem [{\citenamefont {Varpula}\ \emph {et~al.}(2017)\citenamefont
  {Varpula}, \citenamefont {Timofeev}, \citenamefont {Shchepetov},
  \citenamefont {Grigoras}, \citenamefont {Hassel}, \citenamefont {Ahopelto},
  \citenamefont {Ylilammi},\ and\ \citenamefont {Prunnila}}]{Varpula2017}%
  \BibitemOpen
  \bibfield  {author} {\bibinfo {author} {\bibfnamefont {A.}~\bibnamefont
  {Varpula}}, \bibinfo {author} {\bibfnamefont {A.~V.}\ \bibnamefont
  {Timofeev}}, \bibinfo {author} {\bibfnamefont {A.}~\bibnamefont
  {Shchepetov}}, \bibinfo {author} {\bibfnamefont {K.}~\bibnamefont
  {Grigoras}}, \bibinfo {author} {\bibfnamefont {J.}~\bibnamefont {Hassel}},
  \bibinfo {author} {\bibfnamefont {J.}~\bibnamefont {Ahopelto}}, \bibinfo
  {author} {\bibfnamefont {M.}~\bibnamefont {Ylilammi}},\ and\ \bibinfo
  {author} {\bibfnamefont {M.}~\bibnamefont {Prunnila}},\ }\bibfield  {title}
  {\bibinfo {title} {Thermoelectric thermal detectors based on ultra-thin
  heavily doped single-crystal silicon membranes},\ }\href@noop {} {\bibfield
  {journal} {\bibinfo  {journal} {Appl. Phys. Lett.}\ }\textbf {\bibinfo
  {volume} {110}},\ \bibinfo {pages} {262101} (\bibinfo {year}
  {2017})}\BibitemShut {NoStop}%
\bibitem [{\citenamefont {Heikkil\"a}\ \emph {et~al.}(2018)\citenamefont
  {Heikkil\"a}, \citenamefont {Ojaj\"arvi}, \citenamefont {Maasilta},
  \citenamefont {Strambini}, \citenamefont {Giazotto},\ and\ \citenamefont
  {Bergeret}}]{Heikkila2018}%
  \BibitemOpen
  \bibfield  {author} {\bibinfo {author} {\bibfnamefont {T.~T.}\ \bibnamefont
  {Heikkil\"a}}, \bibinfo {author} {\bibfnamefont {R.}~\bibnamefont
  {Ojaj\"arvi}}, \bibinfo {author} {\bibfnamefont {I.~J.}\ \bibnamefont
  {Maasilta}}, \bibinfo {author} {\bibfnamefont {E.}~\bibnamefont {Strambini}},
  \bibinfo {author} {\bibfnamefont {F.}~\bibnamefont {Giazotto}},\ and\
  \bibinfo {author} {\bibfnamefont {F.~S.}\ \bibnamefont {Bergeret}},\
  }\bibfield  {title} {\bibinfo {title} {Thermoelectric radiation detector
  based on superconductor-ferromagnet systems},\ }\href@noop {} {\bibfield
  {journal} {\bibinfo  {journal} {Phys. Rev. Applied}\ }\textbf {\bibinfo
  {volume} {10}},\ \bibinfo {pages} {034053} (\bibinfo {year}
  {2018})}\BibitemShut {NoStop}%
\bibitem [{\citenamefont {Geng}\ \emph {et~al.}(2020)\citenamefont {Geng},
  \citenamefont {Helenius}, \citenamefont {Heikkilä},\ and\ \citenamefont
  {Maasilta}}]{Geng2020}%
  \BibitemOpen
  \bibfield  {author} {\bibinfo {author} {\bibfnamefont {Z.}~\bibnamefont
  {Geng}}, \bibinfo {author} {\bibfnamefont {A.~P.}\ \bibnamefont {Helenius}},
  \bibinfo {author} {\bibfnamefont {T.~T.}\ \bibnamefont {Heikkilä}},\ and\
  \bibinfo {author} {\bibfnamefont {I.~J.}\ \bibnamefont {Maasilta}},\
  }\bibfield  {title} {\bibinfo {title} {Superconductor-ferromagnet tunnel
  junction thermoelectric bolometer and calorimeter with a squid readout},\
  }\href@noop {} {\bibfield  {journal} {\bibinfo  {journal} {J. Low Temp.
  Phys.}\ }\textbf {\bibinfo {volume} {199}},\ \bibinfo {pages} {585} (\bibinfo
  {year} {2020})}\BibitemShut {NoStop}%
\bibitem [{\citenamefont {Varpula}\ \emph {et~al.}(2021)\citenamefont
  {Varpula}, \citenamefont {Tappura}, \citenamefont {Tiira}, \citenamefont
  {Grigoras}, \citenamefont {Kilpi}, \citenamefont {Sovanto}, \citenamefont
  {Ahopelto},\ and\ \citenamefont {Prunnila}}]{Varpula2021}%
  \BibitemOpen
  \bibfield  {author} {\bibinfo {author} {\bibfnamefont {A.}~\bibnamefont
  {Varpula}}, \bibinfo {author} {\bibfnamefont {K.}~\bibnamefont {Tappura}},
  \bibinfo {author} {\bibfnamefont {J.}~\bibnamefont {Tiira}}, \bibinfo
  {author} {\bibfnamefont {K.}~\bibnamefont {Grigoras}}, \bibinfo {author}
  {\bibfnamefont {O.-P.}\ \bibnamefont {Kilpi}}, \bibinfo {author}
  {\bibfnamefont {K.}~\bibnamefont {Sovanto}}, \bibinfo {author} {\bibfnamefont
  {J.}~\bibnamefont {Ahopelto}},\ and\ \bibinfo {author} {\bibfnamefont
  {M.}~\bibnamefont {Prunnila}},\ }\bibfield  {title} {\bibinfo {title}
  {Nano-thermoelectric infrared bolometers},\ }\href@noop {} {\bibfield
  {journal} {\bibinfo  {journal} {APL Photonics}\ }\textbf {\bibinfo {volume}
  {6}},\ \bibinfo {pages} {036111} (\bibinfo {year} {2021})}\BibitemShut
  {NoStop}%
\end{thebibliography}%
\end{document}